\documentclass[conference]{IEEEtran}
\usepackage{cite}
\usepackage{amsmath,amssymb,amsfonts}
\usepackage{algorithmic}
\usepackage{graphicx}
\usepackage{textcomp}
\usepackage{xcolor}
\usepackage{tabularx}
\def\BibTeX{{\rm B\kern-.05em{\sc i\kern-.025em b}\kern-.08em
    T\kern-.1667em\lower.7ex\hbox{E}\kern-.125emX}}
\begin{document}

\title{COVID-19 and Digital Transformation - Developing an Open Experimental Testbed for Sustainable and Innovative Environments (ETSIE) using Fuzzy Cognitive Maps\\
}

\author{\IEEEauthorblockN{1\textsuperscript{st} Wolfgang H{\"o}hl}
\IEEEauthorblockA{\textit{Department of Informatics} \\
\textit{Technical University of Munich (TUM)}\\
Munich, Germany \\
wolfgang.hoehl@tum.de}
}
\maketitle

\begin{abstract}
This paper sketches a new approach using Fuzzy Cognitive Maps (FCMs) to operably map and simulate digital transformation in architecture and urban planning. Today these processes are poorly understood. Many current studies on digital transformation are only treating questions of economic efficiency. Sustainability and social impact only play a minor role. Decisive definitions, concepts and terms stay unclear. Therefore this paper develops an open experimental testbed for sustainable and innovative environments (ETSIE) for three different digital transformation scenarios using FCMs. A traditional growth-oriented scenario, a COVID-19 scenario and an innovative and sustainable COVID-19 scenario are modeled and tested. All three scenarios have the same number of components, connections and the same driver components. Only the initial state vectors are different and the internal correlations are weighted differently. This allows for comparing all three scenarios on an equal basis. The mental modeler software is used (Gray et al. 2013). This paper presents one of the first applications of FCMs in the context of digital transformation. It is shown, that the traditional growth-oriented scenario is structurally very similar to the current COVID-19 scenario. The current pandemic is able to accelerate digital transformation to a certain extent. But the pandemic does not guarantee for a distinct sustainable and innovative future development. Only by changing the initial state vectors and the weights of the connections an innovative and sustainable turnaround in a third scenario becomes possible.
\end{abstract}


\begin{figure*}[htbp]
\begin{minipage}[b]{1.0\textwidth}
\centerline{\includegraphics[width=1.0\textwidth]{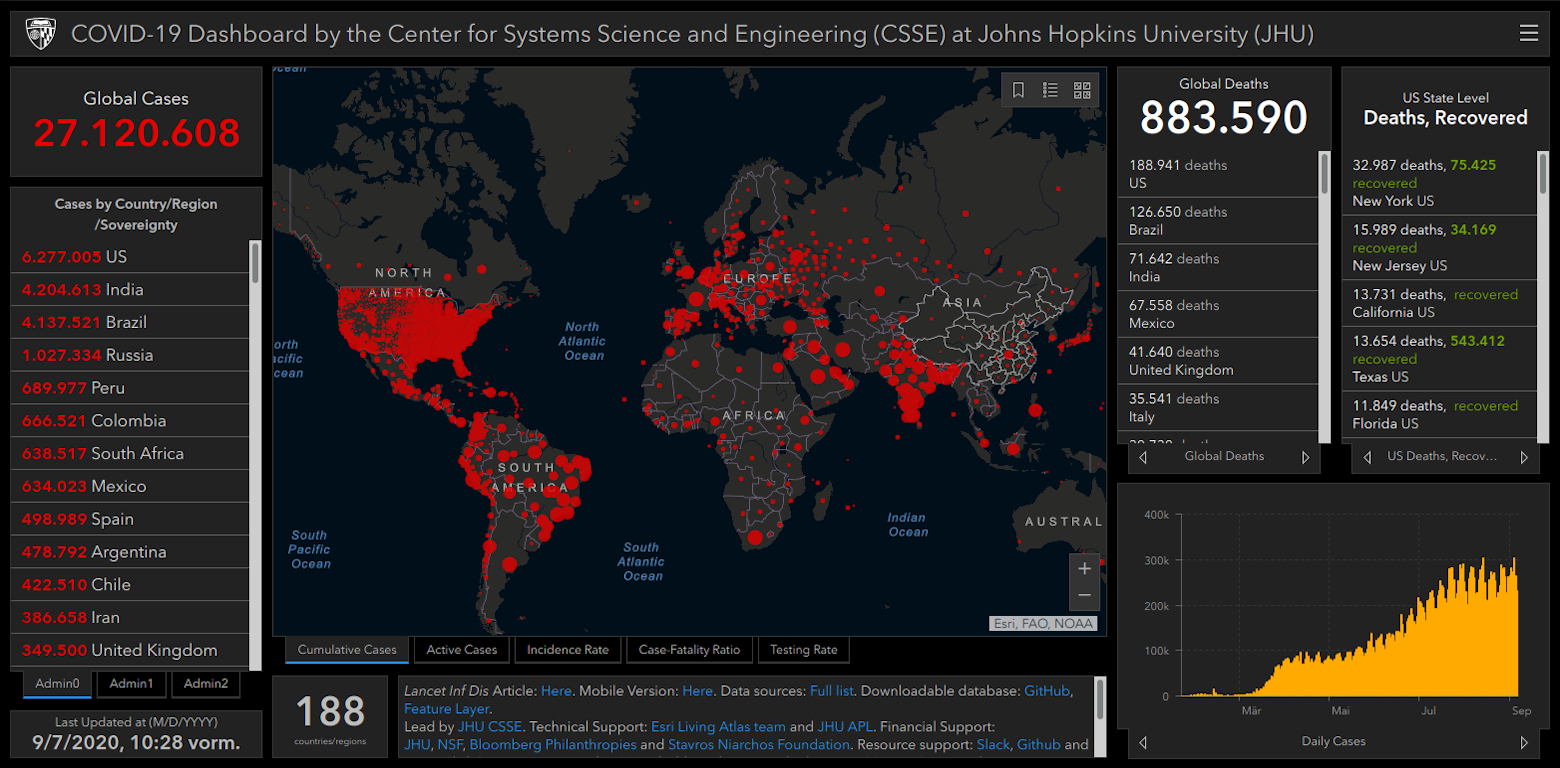}}
\caption{Coronavirus COVID-19 Global Cases by the Center for Systems Science and Engineering (CSSE) at Johns Hopkins University (JHU)}
\label{fig1}
\end{minipage}
\end{figure*}

\section{Introduction}
You probably all know this interactive graphic very well. Figure \ref{fig1}. shows a real-time visualization of the worldwide spread of the corona virus by the Johns Hopkins University in Baltimore \cite{b1}. This interactive map was created with ArcGIS Online. As an architect and planner you probably know similar GIS software that is used in comparable cases. This graphic is a good example for emerging technologies in the current digital transformation.
The current pandemic is accelerating digital transformation. This is not only the opinion of German Federal Minister Andreas Scheuer \cite{b2}. The President of the German Conference of Ministers of Education and Cultural Affairs, Dr. Stefanie Hubig, also expects a boost for digital transformation in schools \cite{b3}. But there are also critical voices. In a petition to the German ministers of education, the Munich teacher Tina Uthoff calls for "an end to distance learning" \cite{b4}. Not all families are able to provide the same level of care and are under great pressure. But there are also other interesting solutions. A Munich wholesale market for gastronomy and retail now supplies food to senior citizens and social food banks with the help of taxi drivers. Logistics software provides the optimized routes for the taxi drivers. The remarkable and decisive factor here is not only using innovative technologies and an innovative idea. Much more it is a specific way to think about solutions, a special "thought style", a guiding mission statement and a clear direction. An effective push always needs a clear direction.
The whole process of digital transformation today is very poorly understood. Therefore this paper sketches a new approach using Fuzzy Cognitive Maps (FCMs) to operably map and simulate the process of digital transformation in architecture and urban planning. Fuzzy Cognitive Maps (FCMs) belong to the so-called soft computing techniques, such as fuzzy logic, neural network theory, genetic algorithms and probabilistic reasoning. Zadeh \cite{b5,b6} defines soft computing as an efficient technique that incorporates human knowledge effectively, deals with imprecision and uncertainty and learns to adapt to unknown or changing environments for better performance. Bonissone \cite{b7} and Jain \cite{b8} mention that soft computing has been successfully applied in many scientific areas such as in engineering, medicine, information systems, business, political and social sciences. FCMs are flexible tools that have been applied in different contexts \cite{b9} including environmental assessment  \cite{b10,b11,b12}, engineering and technological management \cite{b13} and energy \cite{b14}.
Papageorgiou and Groumpos \cite{b15} especially consider FCMs to be capable to deal with situations where the human reasoning process includes uncertain descriptions. FCMs are usually used for modelling specialist knowledge. For data gathering and model building expert interviews and expert opinions are used. Also the delphi method or similar methods can be applied in this initial stage. Secondly FCMs include important modeling means for describing particular domains showing the concepts (variables) and the relationships (connections) between them. For modeling and simulation usually specialized FCM software is used. FCMs are described as a well-established artificial intelligence technique that incorporates ideas from artificial neural networks and fuzzy logic. Thus they are an attractive modeling approach that encompasses advantageous features. On FCMs the following weaknesses can be detected: the main deficiencies of FCMs are the critical dependence on the initial expert's beliefs, the recalculation of the weights corresponding to each concept every time a new strategy is adopted and the potential convergence to undesired equilibrium states. In order to update the initial knowledge of human experts and to combine the human experts' structural knowledge with training from data, specific automated learning methodologies for FCMs my be applied.
This paper presents one of the first applications of FCMs in the context of digital transformation. In the current discussion, a fundamental clarification of the terms and their correlations in digital transformation seems to make sense. Thus this paper provides an overview of literature. In this fundamental discourse, fourteen up-to-date studies on digital transformation in architecture and urban planning  \cite{b17,b18,b19,b20,b21,b22,b23,b24,b25,b26,b27,b28,b29,b30} and twelve studies from the fields of computer science, philosophy of science and media ethics have been searched for stakeholders, specific terms or concepts and correlations in digital transformation \cite{b31,b32,b33,b34,b35,b36,b37,b38,b39,b40,b41,b42}. Terms are clarified, technologies and trends can be identified. The specific "thought styles" and paradigms of the stakeholders are discovered. These elements are then operably mapped and represented in a suitable, integrated model using FCMs. A value added open experimental testbed for sustainable and innovative environments (ETSIE) is developed. Three different digital transformation scenarios are designed. A traditional growth-oriented scenario, a COVID-19 scenario and an innovative and sustainable COVID-19 scenario are modeled and tested. All three scenarios have the same number of components, connections and the same driver components. Only the initial state vectors are different and the internal correlations are weighted differently. This allows for comparing all three scenarios on an equal basis. The mental modeler software is used \cite{b16}. This paper develops the value added ETSIE model as a general framework using FCMs, provides a discussion integrating an overview of literature and highlights directions for future inquiry, but it is not expected to collect extensive empirical data. It is shown, that the traditional growth-oriented scenario is structurally very similar to the current COVID-19 scenario. The current pandemic is able to accelerate digital transformation to a certain extent. But the pandemic does not guarantee for a distinct sustainable and innovative future development. Only by changing the initial state vectors and the weights of the connections an innovative and sustainable turnaround in a third scenario becomes possible.

\newpage
\section{Literature Review}

This chapter presents a fundamental discourse, fourteen up-to-date studies on digital transformation in architecture and urban planning \cite{b17,b18,b19,b20,b21,b22,b23,b24,b25,b26,b27,b28,b29,b30} and twelve studies from the fields of computer science, philosophy of science and media ethics \cite{b31,b32,b33,b34,b35,b36,b37,b38,b39,b40,b41,b42} have been searched for stakeholders, causes and key drivers, technologies and trends in digital transformation. Terms or concepts are clarified, technologies, trends and five indicators of sustainable development could be identified. Specific "thought styles" and paradigms of the stakeholders are discovered.

\subsection{COVID-19 as an Accelerator of Digital Transformation?}

According to Horx [17], Widmann [18], Detting [19] and von der Gracht [20] there are two general characteristics of the corona crisis: on the one hand the novelty of the corona virus and on the other hand the relatively slow course of the crisis compared to other natural disasters. Horx sees the corona crisis as " ... a general slowdown of our world culture", which affects both globalization and our forms of communication. In his opinion, this slowdown will continue after the crisis. "If you take stock, this crisis is a general slowdown of our world culture. This applies to globalization and to our forms of communication. It is a slowdown that will remain [17]." On the contrary for Widmann the corona crisis is not a natural disaster in the conventional sense [18]. According to all we know today, this pandemic is of natural origin. However, pandemics distinguish between people and material assets, and they always affect the socially weak most severely. However, the boundaries to natural disaster remain blurred. Detting outlines a possible more resilient and robust future for society between the polar risks of total interstate and social isolation and the chance of a new sustainable and glocal social-ecological market economy [19]. Depending on the duration of the crisis, von der Gracht sketches four polar scenarios between very moderate and very drastic economic and social consequences [20]. The strong isolation, social conflicts and the virtualization of many areas of life face a rapid recovery and a return to what he calls "normal operations". A positive outlook for a new and innovative social scenario is unfortunately missing here.
COVID-19 is novel and cannot be fought with conventional means. Innovative ideas and technologies must be developed. Due to economic interests, a certain time pressure arises, which pushes the research and development of new technologies. So COVID-19 is an accelerator of digitalization and technological development, in the same sense as other natural disasters can be. Thus the corona crisis can be described as a natural disaster in slow motion. Its relatively slow or time-stretched course compared to other natural disasters allows for a coordinated, reasonable and gradual technological response. The associated deceleration and slowing down of everyday life during the first corona lockdown is certainly still well remembered by us all. The reduction of many growth-driven processes during the Corona crisis thus also led to a certain social and ecological sustainability.

\subsection{Stakeholders of the Digital Transformation}

All industries are directly affected by digital transformation. The processes only differ in speed and intensity. Thierstein describes digital transformation as global and all-encompassing: "Digital transformation permeates all ways of life ... [21]." Goger et al. describe digital transformation as a cross-cutting issue that cuts across all areas of society [22]. The industries only have a different digitization speed. Roland Berger believe that the construction industry is less affected by the corona crisis, as are mechanical engineering and pharmaceutical and medical technology [23]. They see most other industries as being heavily affected, such as airlines, tourism and travel, trade, financing, oil and gas, automotive and logistics. After examining the aforementioned available literature [8-33] four different stakeholders groups can be identified: (P) politics, (R) research and development, (E) economy and (S) civil society as shown in table \ref{tab1}.

\begin{table}[htbp]
\caption{Groups of stakeholders}
\begin{center}
\begin{tabularx}{1.0\columnwidth}{|p{0.02\textwidth}|p{0.15\textwidth}|X|}
\hline
\textbf{} & \textbf{Groups} & \textbf{Description}\\
\hline
{P} & {Politics} & {Cities, districts and communities, federal, state and local authorities} \\
\hline
{R} & {Research and Development} & {Research institutions} \\
\hline
{E} & {Economy} & {Companies, architects and urban planners} \\
\hline
{S} & {Civil Society} & {Population and employees} \\
\hline
\end{tabularx}
\label{tab1}
\end{center}
\end{table}

\subsection {Causes and Drivers of the Digital Transformation}
A study by IE.F and Roland Berger defines following drivers for a successful digital transformation as shown in table \ref{tab2}\cite{b24}. These drivers can be easily assigned to the four different stakeholders groups.

\begin{table}[htbp]
\caption{Groups of stakeholders and assigned causes and drivers
according to IE.F and Roland Berger\cite{b24}}
\begin{center}
\begin{tabularx}{1.0\columnwidth}{|p{0.02\textwidth}|p{0.15\textwidth}|X|}
\hline
\textbf{} & \textbf{Groups} & \textbf{Causes and Key Drivers}\\
\hline
{P} & {Politics} & {Sufficient funding and better political coordination (governance)} \\
{} & {} & {} \\
{} & {} & {Extension of the digital infrastructure} \\
{} & {} & {} \\
{} & {} & {Avoidance of data monopolies and better data protection (data ethics)} \\
\hline
{R} & {Research and Development} & {-} \\
\hline
{E} & {Economy} & {A new business and corporate culture} \\
\hline
{S} & {Civil Society} & {Greater digital literacy} \\
\hline
\end{tabularx}
\label{tab2}
\end{center}
\end{table}

Goger et al. mention following components as key drivers of digital transformation in architecture and urban planning as shown in table \ref{tab3}\cite{b22}:

\begin{table}[htbp]
\caption{Groups of stakeholders and assigned causes and drivers
according to Goger et al.\cite{b22}}
\begin{center}
\begin{tabularx}{1.0\columnwidth}{|p{0.02\textwidth}|p{0.15\textwidth}|X|}
\hline
\textbf{} & \textbf{Groups} & \textbf{Causes and Key Drivers}\\
\hline
{P} & {Politics} & {Sustainability} \\
\hline
{R} & {Research and Development} & {-} \\
\hline
{E} & {Economy} & {Globalization} \\
{} & {} & {Sustainability} \\
{} & {} & {Efficiency and process optimization} \\
\hline
{S} & {Civil Society} & {Urbanization} \\
{} & {} & {Demographic change} \\
{} & {} & {Mobility} \\
{} & {} & {Individualization of work and life models} \\
\hline
\end{tabularx}
\label{tab3}
\end{center}
\end{table}

According to a study by the "Münchner Kreis", product and process quality (79\%) and qualified employees training (78\%) are currently considered the most important success factors in future industrial manufacturing\cite{b24}. Alain Thierstein also mentions the importance of the structure of creative and productive processes. He emphasizes the technological competence of the individual (digital literacy), which is necessary to take responsibility, to evaluate situations and to acquire the new digital environment\cite{b21}. Table \ref{tab4} shows the collected causes and drivers from the these two references:

\begin{table}[htbp]
\caption{Groups of stakeholders and assigned causes and drivers
according to the "Münchner Kreis" \cite{b24} and Thierstein \cite{b21}}
\begin{center}
\begin{tabularx}{1.0\columnwidth}{|p{0.02\textwidth}|p{0.15\textwidth}|X|}
\hline
\textbf{} & \textbf{Groups} & \textbf{Causes and Key Drivers}\\
\hline
{P} & {Politics} & {-} \\
\hline
{R} & {Research and Development} & {-} \\
\hline
{E} & {Economy} & {Product and process quality} \\
\hline
{S} & {Civil Society} & {Qualified employees training} \\
{} & {} & {Digital literacy} \\
\hline
\end{tabularx}
\label{tab4}
\end{center}
\end{table}

\subsection{Degree of Automation and Digitization Speed as System Indicators}

Roland Berger see digitization in the workplace and the home office as a direct effect of increasing digital transformation\cite{b23}. Thus Berger confirm the interconnections between the use of digital technology, user behavior and mobility (e.g. home office, digital collaboration). Goger et al. estimate that there is a 59\% probability of substituting construction occupations with Industry 4.0. At least they call it "degree of automation" or "probability of automation" and present their "vision of digital construction"\cite{b22}. During the early phases of development, better variant studies for decision making become possible through visualization and simulation, better and significantly more transparent information exchange. Central digital twins of the building and applications of BIM in integral planning are getting in the field of vision, as well as processes for automated quality assurance and compliance with building standards, simplified tendering on a digital basis and fully automated reading of masses and quantities. During construction, improved logistics through RFID tracking with location allocation would be possible. Digital recording of delivery bills and material parameters and a complete documentation would digitally supplement the construction progress. Simplified surveys could be provided through drone flights. Innovative manufacturing processes could support or even replace conventional procedures. IoT and Big Data will be deployed during building use for automatic building data collection, automatic ventilation and building climate control. During demolition, the building can serve as a raw material store (urban mining). Through a digitally transparent process, all quantities and materials can be known in advance.
Katz et al. describe digitization speed as a composite factor of affordability, infrastructure investment, network access, capacity, usage and human capital\cite{b25}. As digitization speed can not be assigned to one stakeholder only, a new category of indicators for system behaviour is created. With the degree of automation and the digitization speed the authors name two important indicators for the social sustainability of digital transformation. Table \ref{tab5} shows the extended list of stakeholders and indicators.

\begin{table}[htbp]
\caption{Groups of stakeholders and indicators with assigned causes and drivers
according to Katz et al.\cite{b25}}
\begin{center}
\begin{tabularx}{1.0\columnwidth}{|p{0.02\textwidth}|p{0.15\textwidth}|X|}
\hline
\textbf{} & \textbf{Groups} & \textbf{Causes and Key Drivers}\\
\hline
{P} & {Politics} & {Infrastructure investment} \\
\hline
{R} & {Research and Development} & {-} \\
\hline
{E} & {Economy} & {-} \\
\hline
{S} & {Civil Society} & {Affordability} \\
{} & {} & {Network Access} \\
{} & {} & {Capacity} \\
{} & {} & {Digital Usage} \\
{} & {} & {Human capital} \\
\hline
{I} & {Indicators} & {Digitization speed} \\
{} & {} & {Degree of Automation} \\
\hline
\end{tabularx}
\label{tab5}
\end{center}
\end{table}

The "vision of digital construction" has a high social relevance in addition to its economic relevance. Schüller developed the term "digital literacy"\cite{b27}. He also names it among the so-called success factors of the digital transformation. The speed of digitization in the respective industries will depend on the digital literacy of those involved in the process and the quality of products and services.
Simondon warns against such purely economically driven automation\cite{b28}. He describes automation by the well-known image of a fully automatic and autonomous robot. Simondon sees in it only an abstract mythical object, without any relevance for a practical and really innovative technological development. "Automatism and its use in the form of the industrial organization called automation has far more of an economic or social than a technical meaning \cite{b28}." In order to ensure a functioning human-machine interaction, Simondon advocates "open machines" with a certain "margin of uncertainty". Developments such as artificial intelligence, algorithms for the individualization and personalization of interfaces or even autonomous driving could easily be assigned to this fully digitalized automation.
According to Baumanns, the interdisciplinary collaboration across all phases of a building's life cycle (design, construction, operation) also still offers great potential for digital transformation. This includes a central, digital building model and digital, interdisciplinary communication and coordination. Baumanns defines the following megatrends\cite{b29} as shown in table \ref{tab6}.As growth drivers the authors describe residential construction, infrastructure and transport. The general construction volume is currently benefiting from low loan and interest rates. As competitive advantages for companies he sees the areas of Smart Home, Smart Building and Building Information Modeling (BIM). As current opportunities he mentions globalization, specialization and expansion of the company's own portfolio in the existing value chain and the increasing use of digital technology. The greatest potential he assumes in the field of logistics, in the digital collection and analysis of data and in the automation of construction work. He names five phases: Logistics, procurement, production, marketing/sales and after sales.

\begin{table}[htbp]
\caption{Groups of stakeholders and indicators with assigned causes and drivers
according to Baumanns\cite{b29}}
\begin{center}
\begin{tabularx}{1.0\columnwidth}{|p{0.02\textwidth}|p{0.15\textwidth}|X|}
\hline
\textbf{} & \textbf{Groups} & \textbf{Causes and Key Drivers}\\
\hline
{P} & {Politics} & {Sustainability} \\
{} & {} & {Low interest rate policy} \\
\hline
{R} & {Research and Development} & {Digitization and Technology} \\
\hline
{E} & {Economy} & {Globalization} \\
{} & {} & {Smart Home / Smart Building} \\
{} & {} & {Building Information Modeling (BIM)} \\
{} & {} & {Specialization} \\
{} & {} & {Extension of the own portfolio (value chain)} \\
\hline
{S} & {Civil Society} & {Urbanization} \\
{} & {} & {Demographic change} \\
\hline
{I} & {Indicators} & {-} \\
\hline
\end{tabularx}
\label{tab6}
\end{center}
\end{table}

\subsection{Technologies, Trends and Primary Energy Consumption}

Analogous to Industry 4.0, Goger et al. coined the term "Construction 4.0" \cite{b22}. The Digital Roadmap Austria identifies ten technology areas with enormous development potential that could also be of significance for architecture and urban planning\cite{b30}. A study by BRZ Deutschland GmbH. names the following six IT-trends in the construction industry\cite{b31}.
However, in these studies there is no distinction between conventional and sustainable technologies. Some of the technologies mentioned, such as block chain technology, are currently heavily criticized, because of their disproportionate energy consumption. The sustainability of IoT is also under current discussion. The integrated IT and the lack of software updates for intelligent household appliances could lead to a shorter lifetime of usual household appliances such as refrigerators. Moreover today there is no clear distinction and recommendation between sustainable technologies and non-sustainable technologies. Additionally there are currently no comparable indicators of the systemic energy efficiency of the individual technologies. This could be ensured by a technology-specific ecological footprint.
To take these factors into account, two further indicators will be introduced to assess the sustainability of the overall system: primary energy consumption and the ratio of sustainable processes and technologies to the total amount of digital processes and technologies as shown in table \ref{tab7}. To my opinion, the "virtual project space" and "cloud computing" can be described more as distinct technologies and less as a trend. Established in science, the term "Collaborative Virtual Environments" (CVE) has long been used to describe these technical solutions. The same applies to BIM. I would only use the term trend in connection with an overarching scenario based on a specific thought style, mindset or paradigm.

\begin{table}[htbp]
\caption{Groups of stakeholders and indicators with assigned technologies and trends
according to Goger et al.\cite{b22}, Digital Roadmap Austria\cite{b30} and BRZ\cite{b31}}
\begin{center}
\begin{tabularx}{1.0\columnwidth}{|p{0.02\textwidth}|p{0.15\textwidth}|X|}
\hline
\textbf{} & \textbf{Groups} & \textbf{Technologies and Trends}\\
\hline
{P} & {Politics} & {Big Data} \\
\hline
{R} & {Research and Development} & {Artificial Intelligence (AI)} \\
\hline
{E} & {Economy} & {Open Innovation} \\
{} & {} & {Intelligent energy networks (smart grids)} \\
{} & {} & {Intelligent materials (4D)} \\
{} & {} & {New manufacturing processes (3D printing)} \\
{} & {} & {Extended Reality (xR)} \\
{} & {} & {5G mobile phone standard} \\
{} & {} & {IoT and cloud computing} \\
{} & {} & {Building Information Modeling (BIM)} \\
{} & {} & {Blockchain Technology} \\
{} & {} & {Virtual Project Space} \\
\hline
{S} & {Civil Society} & {Mobility} \\
{} & {} & {Social security} \\
{} & {} & {IT Security} \\
\hline
{I} & {Indicators} & {Primary Energy Consumption} \\
{} & {} & {} \\
{} & {} & {Total amount of digital processes and technologies} \\
{} & {} & {} \\
{} & {} & {Sustainable digital processes and technologies} \\
\hline
\end{tabularx}
\label{tab7}
\end{center}
\end{table}

\subsection{Digitization, Digital Transformation, Innovation and Sustainability}

\textit{F1. Digitization, Digital Transformation or Innovation?}

Many authors mention digitization itself as a driver for major changes in work processes. It would be much more accurate to name technological change through research and development as a driver of digitization. Thus research and development appeared as a stakeholder on the scene. The newly developed technology and the associated research and development (not digitization or digital transformation) are the drivers. To a large extent new technological developments lead to changes in work processes. Digital transformation can be better described as the systemic end result or as a significant scenario in this macrosocietal transformation.
Digitization and technology are also listed as trends in many studies. What is missing here is a precise distinction between digitization (or digital transformation) as a cross-system scenario on the one hand and digital technology as a subordinate systemic element resulting from research and development on the other. Just as BIM, smart homes and smart buildings can be better described as technologies rather than as trends.
Digitization and digital transformation are not used as congruent terms in this paper. Digital transformation better desribes the whole process. While digitization can usually be understood as the "conversion of analogue quantities into digital ones". But neither digitization nor digital transformation does not necessarily mean innovation. The digitization of an existing process does not necessarily have to be innovative per se. Purely economic and efficiency-oriented digitization is the adaptation of an existing process to modern technology. However, this does not necessarily make the process new and certainly not sustainable. Innovation always means a previously unknown, new solution to a problem. The innovative idea is also fundamental to innovation, which then leads to a new solution by recombining existing resources. At the beginning of an innovation there is always an idea and a new mindset, a new way of thinking about things. The implementation into a suitable technological solution is subordinated.\\

\textit{F2. Sustainability}

In the examined literature sustainability is often used as an ambiguous term or concept. For example Baumanns cites sustainability itself as a current megatrend\cite{b29}. Thus sustainability appears in a completely different context than in IE.F and Roland Berger\cite{b24} and Goger et al.\cite{b22}. There sustainability is mentioned as a trend. In many studies sustainability is seen as a way of thinking, a thought style or a paradigm. It can also be interpreted as a social factor, such as a sustainable lifestyle. Or it can be interpreted as an economic factor such as a new sustainable business and corporate culture. Sustainability is often mentioned in connection with an economically efficiency-oriented process optimization. Many experts also expect sustainability to have positive effects on future energy and resource consumption. Last but not least it could stand for either a sustainable research paradigm or a sustainable political "thought style". The term sustainability can also be interpreted in a political sense as the amount of political actions aiming climate protection. Thus sustainability itself cannot precisely be assigned to one stakeholders group only. The term finds correspondences in all four stakeholders groups as a "thought style" or a specific way of thinking. So it will reappear again as a political "thought style", a "research paradigm", a public "thought style" and an "economic business and corporate culture". It also can be defined as another seperate concept, such as political or economical actions for climate protection. Sustainability also can be seen as a quality of the whole system behaviour, when we think of technologies and trends. So indicators such as  primary energy consumption, the degree of automation, digitization speed and the ratio of sustainable processes and technologies to the total amount of digital processes and technologies can be seen strongly related to the main concept of sustainability as well.

\subsection{Thought Styles and Paradigms as Individual Mindsets for Digital Transformation}

In the aforementioned sections another group of causes and drivers of the digital transformation emerges. It is the group of thought styles, paradigms, ways of thinking or reasoning. This group will be explained in more detail in the next section.
Ian Hacking coined the term "style of reasoning" in connection with the continuous change in mission statements \cite{b32}. He essentially followed Alistair Crombie (1915-1996), who claimed that there are different scientific methods of knowledge that have emerged in certain areas of human history. Ludwik Fleck developed the analogous concept of "thinking styles"\cite{b33}. Rudolf S. Kuhn uses the term "paradigm shift" to describe the change in basic conditions for theory formation in science, such as concept formation, methods of observation and technology used \cite{b34}. Nowadays, terms such as "theory dynamics" and "theory change" are often used to describe these phenomena. Luca Sciortino\cite{b35} also mentions Michel Foucault's 'episteme'\cite{b36} and the "research program" of Imre Lakatos\cite{b37}. All these authors thus describe phenomena in connection with change on the basis of changed mindsets and models, ways of thinking and paradigms. These new perspectives lead to a new basis, enable new methods and new technologies and, last but not least, lead to real innovation.
Mission statements usually describe a desired goal or an ideal state. They are at home in many areas. They are often used in corporate culture. A mission statement serves as orientation and motivation for the employees of a company, possibly provides information about the product range and activities of a company, can provide a certain framework for the public appearance of a company and even influences the market value of a company. Kühl distinguishes between three sides of a company\cite{b38}: the so-called "show side", the "formal side" and the "informal side". The front side shows the external image, the facade of the company. The so-called formal side forms the official set of rules for all employees. The informal side describes Kühl as the sum of "... ingrained practices and ways of thinking, deviations from official rules and from cultivated myths, dogmas and fictions."
But there are also certain mindsets or paradigms for technical or social developments. They are often closely related to current research and development and formulate certain utopias in these fields. They are ideal models or wishful thinking about future possibilities of technology and society. These models and utopias are often the initial impetus, also for new research projects.
Some of the guiding paradigms for digital transformation date back to the 1990s. With "ubiquitous computing" Mark Weiser described an omnipresent mobile use of spatial information accessible to everyone, without visible interfaces and end devices\cite{b39}. Neil Gross expected "that in the future spontaneous computer networks will emerge and form a 'giant digital creature'\cite{b40}. He thus describes the current Internet of Things (IoT). "Pervasive Computing" and "Ambient Intelligence" also describe related topics today, but with different orientations\cite{b41}. Ambient Intelligence deals with intelligent systems embedded in the environment that support the user in his activities. In contrast to purely commercial considerations, however, the focus here is often on social and procedural issues. Smart Cities" and "Smart Homes" also belong to these thematic fields. The most important features of ubiquitous computing are the disappearance of hardware and user interfaces, the adaptivity and self-organization of the digital system, automatic context perception, ubiquitous availability of information, and global and local connectivity. I fondly remember a photo of Archigram from the sixties. It shows a telephone in a lying tree trunk, somewhere out in nature.
Technical and social utopias are always intertwined. Social models are not always easy to grasp. Tanner describes among other things the concept of "common sense"\cite{b42}. He understands this to mean social rules and behavior, identity-forming myths and stories, the knowledge of things that "one" does or does not do. "Common sense is the ability to think logically without using specialized or advanced knowledge\cite{b42}." The digital transformation in a pluralistic society therefore does not only know one single mindset or paradigm. Companies have their own mission statements, research and development work according to their own ways of thinking and standards, and the population pursues its own individualized life and work ideas, as sometimes, among other things, a sustainable lifestyle. Nevertheless, these mission statements remain powerful drivers for our everyday life.

\subsection{Clarifying terms, concepts and indicators}

Table \ref{tab8} summarizes all clarified concepts and terms from the former literature examination. We can state four stakeholders (politics, research and development, economy and civil society) such as mentioned in table \ref{tab1}. Sustainability shows up in three different correlations such as in table \ref{tab7}. First it can be a "thought style" for all four stakeholders (P1,  R1, E1, S1) ("a new business and corporate culture" in table \ref{tab2}, "sustainability" and "individualization of work and life models" in table \ref{tab3}. Secondly it can include political actions for climate protection (P2). Third sustainability can be related to the indicators of primary energy consumption (I1) and the ratio of sustainable processes and technologies (I5) to the total amount of digital processes and technologies (I4) as mentioned in table 7. These indicators can be completed by the degree of automation (I2) and the digitization speed (I3) as mentioned in table \ref{tab5}. So we are able to define a group of five indicators finally. 
The terms of infrastructure investment (P3) ("extension of the digital infrastructure"), financing and coordination (P4) ("governance", "low interest rate policy") and prevention of data monopolies (P5) ("data ethics" or "avoidance of data monopolies and better data protection" or "IT security") were mentioned in tables \ref{tab2}, \ref{tab5}, \ref{tab6} and \ref{tab7}. The terms of "demographic change" and "urbanization" as found in tables \ref{tab3} and \ref{tab6} can be expressed in population figures. So both are summarized in the concept of population (S2). Digital literacy (S3) and education (P6) were added as social and political concepts, such as mentioned in table \ref{tab2} ("greater digital literacy"), table \ref{tab4} ("digital literacy" and "qualified employees training") and table \ref{tab5} ("human capital"). Research and Development (R2) was introcuded to reflect the terms of "digitization and technology" and terms of technological development and trends such as in tables \ref{tab6} and \ref{tab7}. Product and process quality (E2) was inserted to reflect the terms of "globalization", "specialization", "extension of the own portfolio", "product and process quality" and "efficiency and process optimization" such as in tables \ref{tab3}, \ref{tab4} and \ref{tab6}. Digital Usage (S4), affordability (S5), network access (S6), capacity (S7) were adopted directly from table \ref{tab5}. Mobility (S8) was also listed as in tables \ref{tab3} and \ref{tab7}.

\begin{table}[htbp]
\caption{Clarified concepts and terms}
\begin{center}
\begin{tabularx}{1.0\columnwidth}{|p{0.02\textwidth}|p{0.15\textwidth}|p{0.02\textwidth}|X|}
\hline
\textbf{} & \textbf{Groups} & \textbf{} & \textbf{Concepts}\\
\hline
{P} & {Politics} & {P1} & {Political "Thought Style"}\\
{} & {} & {P2} & {Climate Protection}\\
{} & {} & {P3} & {Infrastructure Investment}\\
{} & {} & {P4} & {Financing and Coordination (Governance)}\\
{} & {} & {P5} & {Prevention of Data Monopolies (Data Ethics)}\\
{} & {} & {P6} & {Education}\\
\hline
{R} & {Research and Development} & {R1} & {Research Paradigm}\\
{} & {} & {R2} & {Research and Development}\\
\hline
{E} & {Economy} & {E1} & {Business and Corporate Culture}\\
{} & {} & {E2} & {Product and Process Quality}\\
\hline
{S} & {Civil Society} & {S1} & {Public "Thought Style"}\\
{} & {} & {S2} & {Population}\\
{} & {} & {S3} & {Digital Literacy}\\
{} & {} & {S4} & {Digital Usage}\\
{} & {} & {S5} & {Affordability}\\
{} & {} & {S6} & {Network Access}\\
{} & {} & {S7} & {Capacity}\\
{} & {} & {S8} & {Mobility}\\
\hline
{I} & {Indicators} & {I1} & {Primary Energy Consumption}\\
{} & {} & {I2} & {Degree of Automation}\\
{} & {} & {I3} & {Digitization Speed}\\
{} & {} & {I4} & {Total amount of digital processes and technologies}\\
{} & {} & {I5} & {Sustainable digital processes and technologies}\\
\hline
\end{tabularx}
\label{tab8}
\end{center}
\end{table}

\newpage
\section{Materials and Methods}

\subsection{Fuzzy Cognitive Maps (FCMs), Definition and Background}

As mentioned in the introduction FCMs are usually used to investigate complex systems. They consist of a network of concepts and weighted interconnections. The technique of FCMs is often deployed to reveal a dynamic system's behaviour, describing how the system could evolve in time through causal relationships and for the evaluation of alternatives. Therefore this approach is considered useful in the context of scenario planning and decision making. For example "Integrated World System Modelling" or an "Integrated Global System Model" is particularly suitable for mapping complex processes using FCMs. The concept of Fuzzy Cognitive Maps (FCMs) has been introduced by Bart Kosko in 1986 \cite{b43}. He suggested their use to those knowledge domains that involve an high degree of uncertainty.  He extended the work of Axelrod, who found the technique of Fuzzy Cognitive Maps (FCMs) representing a natural extension of cognitive maps by embedding to them the use of Fuzzy Logic \cite{b44}. He also introduced cognitive maps in the context of decision making for the representation of social scientific knowledge and decision making processes in the field of social and political systems. Mention has to be made to the fact that the cognitive mapping approach to decision making uses elements from other fields, such as psychology and graph theory. Since then, FCMs have proven to be a useful method for the systematic collection of knowledge and for the graphical representation of causal relationships and are established in many "hard" and "soft" sciences. Caselles names several research groups that are currently working on global models to simulate the consequences of scenarios and intervention strategies in the world system \cite{b45}. As prominent examples he mentions the "Regional Earth System" of the Earth System Science Interdisciplinary Centre at the University of Maryland, the "Millennium Project" of the World Bank, the "Integrated Global System Model" of MIT and the "Australian Stocks and Flows Framework (ASFF)". Bottero et al. for example, successfully use FCMs to model urban resilience dynamics and to support scenario planning and strategic decision making \cite{b46}.
The procedure follows the conventional method of a simulation. Cloud describes the usual professional development of model-building and simulation proceeds in three stages \cite{b47}: Theoretical basis and monitoring (observation), problem and solution attemts, criticism and elimination of possible errors and mistakes. In the first step you are confronted with a kind of "source system". This system is real and can be observed or monitored experimentally. Let us assume this is the process of digital transformation in architecture and urban planning. For monitoring and data gathering usually expert interviews, expert opinions or literature reviews are used. Also the delphi method or similar methods can be applied in this initial stage. From there we can derive our input values. The second step is building an abstract model from that source system (idealization and abstraction). Usually only one specific aspect is modeled to clarify one specific question. From the abstract model we can derive an applicable or operable model, for example in a computer simulation. Now we could run an iterative process to test the model and to generate certain scenarios. The output values can be validated against the source system, errors and mistakes can be eliminated and we can start the process again from the beginning. From there we can start an optimization loop. Simulations usually are conducted within a large parameter space to simulate many possible situations. To optimize the target size, the input variables must be varied. This can be done by try-and-error, but this takes a lot of time and effort. Therefore, in the last years, evolutionary optimization methods (e.g. data farming) have been developed to reach the whole realm of a simulation's test bed. These solutions usually are performed in iterative steps to find an optimal solution. Usually this is done by High Performance Computer (HPC) environments. The results can be depicted in scenarios. The procedure can be repeated until an optimized state is reached.
Today, independent learning algorithms help to optimize various system-internal parameters, such as the weighting of correlations. Different learning paradigms and software packages can be distinguished. Felix et al. \cite{b48} name seven software applications available today: FCM Modeler, FCM Designer, FCM Tool, JFCM, Mental Modeler, ISEMK and FCM Expert. Because of the participatory, web-based solution and because of the ease of use without programming knowledge, this paper uses the Mental Modeler software, developed by Gray et al. \cite{b16}.

\subsection{Properties of FCMs}

\begin{itemize}
\item Concepts or components: C1, C2, ... Cn are the system-constituting variables;
\item State vector: A = (a1, a2, ... an) where ai is the initial state and the value of the general term Ci. The values assigned to the terms are usually in the range [0;1];
\item Directed edges: they symbolize the causality between the terms C1, C2 and are represented as arrows or double arrows;
\item Adjacency matrix: E = \{eij\}, where eij is the weight (w) of the directed edge CiCj. The values assigned to each relation are in the range [-1;1]. The value 0 means, that there is no causal relation between the terms Ci and Cj.\\
\end{itemize}

\begin{figure}[htbp]
\centerline{\includegraphics[width=1.0\columnwidth]{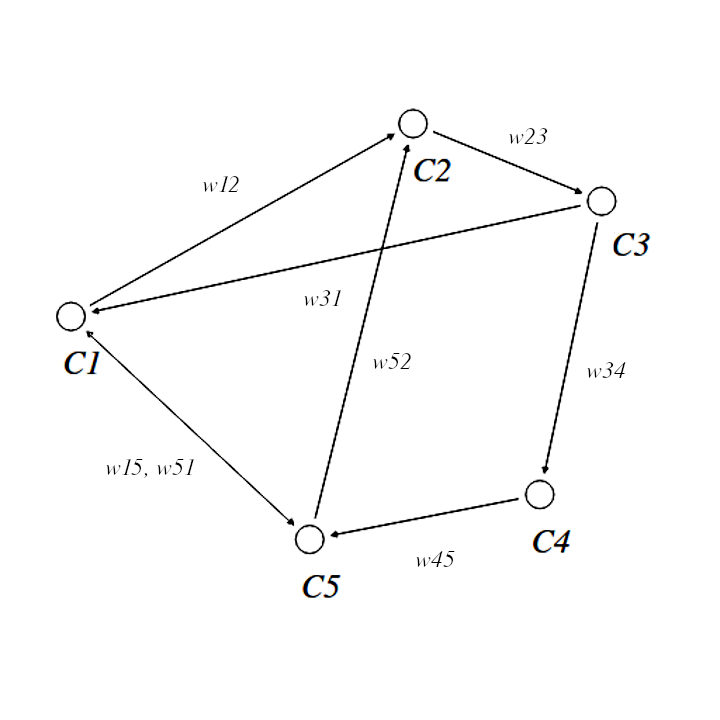}}
\caption{Graphical Representation of a Fuzzy Cognitive Map (FCM)}
\label{fig2}
\end{figure}

FCMs can be described by different kinds of representation: On the one hand by a graphical representation, the concepts or components and the directed edges. On the other hand by a mathematical representation, which consists of a table mentioning all state vectors in an adjacency matrix. Figure \ref{fig2}. shows a graphical representation of a Fuzzy Cognitive Map (FCM). Table \ref{tab9} shows the corresponding adjacency matrix.

\begin{table}[htbp]
\caption{Adjacency Matrix of a Fuzzy Cognitive Map (FCM)}
\begin{center}
\begin{tabularx}{1.0\columnwidth}{|X|X|X|X|X|X|}
\hline
{} & {C1} & {C2} & {C3} & {C4} & {C5}\\
\hline
{C1} & {0} & {1} & {0} & {0} & {1}\\
\hline
{C2} & {0} & {0} & {1} & {0} & {0}\\
\hline
{C3} & {1} & {0} & {0} & {1} & {0}\\
\hline
{C4} & {0} & {0} & {0} & {0} & {1}\\
\hline
{C5} & {1} & {1} & {0} & {0} & {0}\\
\hline
\end{tabularx}
\label{tab9}
\end{center}
\end{table}

\subsection{Developing an Open Experimental Testbed for Sustainable and Innovative Environments (ETSIE) using Mental Modeler}

One of the great minds of simulation and statistics, George Edward Pelham Box used to say: "All models are wrong, but some of them are useful."\cite{b49} This paper presents one of the first applications of FCMs in the context of digital transformation. It aims to design a fuzzy but useful general framework and proof of concept for future inquiry. In this paper terms and concepts for model building are collected from a literature review. Using this "fuzzy" method it is also common for experts to define the correlations in the model. Thus the correlations and causalities correspond to reasonable assumptions by the author. For simulation the mental modeler software is used, developed by Gray et al.\cite{b16}. FCMs can describe and simulate the dynamic behavior of systems. Simulating the system behaviour is usually based on mathematical operations and takes place in iterative steps. Unfortunately, the Mental Modeler software is very limited in this respect and does also not allow for the use of dedicated learning algorithms. However, scenario building is very simple, user-friendly and allows for a quick visualization of results. Therefore, this paper only uses the given possibilities of scenario building in the Mental Modeler Software and does not include an additional, algorithmically supported simulation in iterative steps.
This paper develops an open experimental testbed for sustainable and innovative environments (ETSIE) as a general framework using FCMs. It highlights directions for future inquiry, but it is not expected to collect extensive empirical data. The ETSIE model is developed in the spirit of Open Innovation. The model itself and all data are openly available to all interested people for review and further research.
From the current literature review it can be observed that the process of digital transformation is often seen in a predominantely growth-oriented economic sense. During the current pandemic it can be observed as well, that sustainability is not necessarily written in capital letters. Developing innovative and sustainable social and economic models is currently not the first priority. Therefore three different digital transformation scenarios are designed. A traditional growth-oriented scenario, a COVID-19 scenario and an innovative and sustainable COVID-19 scenario are modeled and tested. All three scenarios have the same number of components, connections and the same driver components. Only the initial state vectors are different and the internal correlations are weighted differently. This allows for comparing all three scenarios on an equal basis. The following three scenarios are illustrated:\\

\begin{itemize}
\item Traditional, growth-oriented scenario
\item COVID-19 scenario (natural disaster)
\item Innovative and sustainable COVID 19 scenario\\
\end{itemize}

Five indicators provide information on the quality of the different scenarios:\\

\begin{itemize}
\item Primary energy consumption (I1)
\item Degree of automation (I2)
\item Digitization speed (I3)
\item Total number of existing digital processes and technologies (I4)
\item Number of sustainable digital processes and technologies (I5)\\
\end{itemize}

Inside the given Mental Modeler software platform data is calculated by the capabilities of the software. For setting the state vectors the software only allows values between -1 and +1. State vectors are set by a reasonable assumption by the author to get significant results. One single scenario is generated at one calculation time. Simulation results are taken directly from the given software. No further iterations, no evolutionary optimization methods or learning algorithms are used. No other than the mentioned scenarios are generated. The three scenarios then subsequently are compared to each other to see, if the ETSIE model delivers useful and significant results and is able to be a proof of concept.

\newpage
\subsection{Stakeholders}

The following four different groups of people can be identified as stakeholders for the ETSIE model:\\

\begin{itemize}
\item (P) Politics (cities, counties and municipalities - federal, state and local)
\item (R) Research and development
\item (E) Economy (companies, architects and urban planners)
\item (S) Civil society (population and employees)\\
\end{itemize}

\subsection{Concepts}

The concepts from table \ref{tab8} are directly transferred to table \ref{tab10}. This table shows the concepts of the ETSIE model with a detailed description. All concepts are assigned to the different stakeholders: politics (Pn), research and development (Rn), economy (En), civil society (Sn) and indicators (In).

\begin{table}[htbp]
\caption{List of concepts}
\begin{center}
\begin{tabularx}{1.0\columnwidth}{|p{0.02\textwidth}|p{0.15\textwidth}|X|}
\hline
\textbf{} & \textbf{Name} & \textbf{Description}\\
\hline
{P1} & {Political Thought Style} & {Common sense or style of reasoning} \\
{P2} & {Climate Protection} & {Political actions for climate protection} \\
{P3} & {Infrastructure Investment} & {Political actions for infrastructure investment} \\
{P4} & {Financing and Coordination (Governance)} & {Political actions regarding subsidies and governmental support} \\
{P5} & {Prevention of Data Monopolies (Data Ethics)} & {Political actions regarding data ethics} \\
{P6} & {Education} & {Political actions for educational support} \\
\hline
{R1} & {Research Paradigm} & {Common sense or style of reasoning} \\
{R2} & {Research and Development} & {Research actions} \\
\hline
{E1} & {Business and Corporate Culture} & {Common sense or style of reasoning} \\
{E2} & {Product and Process Quality} & {Economical actions regarding product and process quality} \\
\hline
{S1} & {Public Thought Style} & {Common sense or style of reasoning} \\
{S2} & {Population} & {Total amount of population} \\
{S3} & {Digital Literacy} & {Amount of digitally educated people} \\
{S4} & {Digital Usage} & {Average daily usage time of digital media} \\
{S5} & {Affordability} & {Average price of digital technology} \\
{S6} & {Network Access} & {Amount of people having digital network acces} \\
{S7} & {Capacity} & {Average available bandwidth per inhabitant} \\
{S8} & {Mobility} & {Average daily usage time of means of transport} \\
\hline
{I1} & {Primary Energy Consumption} & {Primary energy demand (Qp)} \\
{I2} & {Degree of Automation} & {Percentage of digital process automation} \\
{I3} & {Digitization Speed} & {Composite factor of population, digital literacy, digital usage, affordability, capacity and network access} \\
{I4} & {Total amount of digital processes and technologies} & {Total amount of digital processes and technologies} \\
{I5} & {Sustainable digital processes and technologies} & {Percentage of sustainable processes and technologies} \\
\hline
\end{tabularx}
\label{tab10}
\end{center}
\end{table}

\subsection{Three Alternative Scenarios}

In the following, three alternative scenarios, including the COVID-19 scenario, are depicted using FCMs in the ETSIE model and evaluated according to the five basic indicators mentioned above. Table \ref{tab11} shows a description of all three alternative scenarios:\\

\begin{itemize}
\item Scenario 1 - Traditional, growth-oriented scenario
\item Scenario 2 - COVID-19 scenario (natural disaster)
\item Scenario 3 - Innovative and sustainable COVID-19 scenario
\end{itemize}

\begin{table}[htbp]
\caption{Three alternative scenarios}
\begin{center}
\begin{tabularx}{1.0\columnwidth}{|p{0.15\textwidth}|X|}
\hline
\textbf{Alternative scenarios} & \textbf{Description}\\
\hline
{(1) Traditional growth-oriented scenario} & {This scenario includes a traditional form of the growth-oriented economy, no political subsidies for sustainable developments, no sustainable research and development, an increase of individual life and work forms and an increase of individual mobility}\\
\hline
{(2) COVID-19 scenario (natural disaster)} & {This scenario is characterized by rising governmental expenses, economical subsidies, growing investment in research, but no dedicated promotion of sustainable technologies}\\
\hline
{(3) Innovative and sustainable COVID-19 scenario} & {This scenario shows an innovative approach towards a sustainable digital transformation under COVID-19 conditions.}\\
\hline
\end{tabularx}
\label{tab11}
\end{center}
\end{table}

All three scenarios are built from the same number of concepts (or components) and connections. They all have the same density and the same number of connections per component. All three models have exactly the same complexity score and show six driver components as listed below:\\

\begin{itemize}
\item P1 - Political Thought Style
\item R1 - Research Paradigm
\item E1 - Business and Corporate Culture)
\item S1 - Public Thought Style
\item S2 - Population
\item S5 - Affordability\\
\end{itemize}

This structural design, the general form of the models, remains the same in all three scenarios. Only the weights of the connections between the nodes (or components) change. This is intended to make it easier and clearer to recognize deviating results and behaviour of the individual models.

\newpage
\section{Results}

\subsection{Scenario 1 - Traditional Growth-Oriented Scenario}

\begin{figure*}[htbp]
\begin{minipage}[b]{1.0\textwidth}
\centerline{\includegraphics[width=1.0\textwidth]{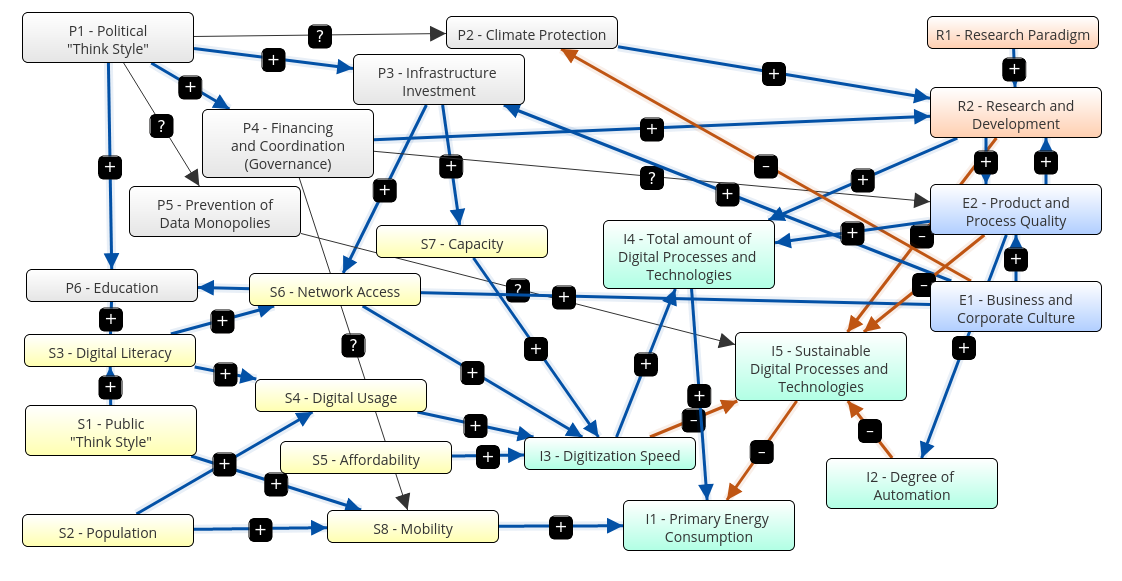}}
\caption{Graphical representation . Traditional growth-oriented scenario (Screenshot from Mental Modeler Software)}
\label{fig3}
\end{minipage}
\end{figure*}

\begin{figure*}[htbp]
\begin{minipage}[b]{1.0\textwidth}
\centerline{\includegraphics[width=1.0\textwidth]{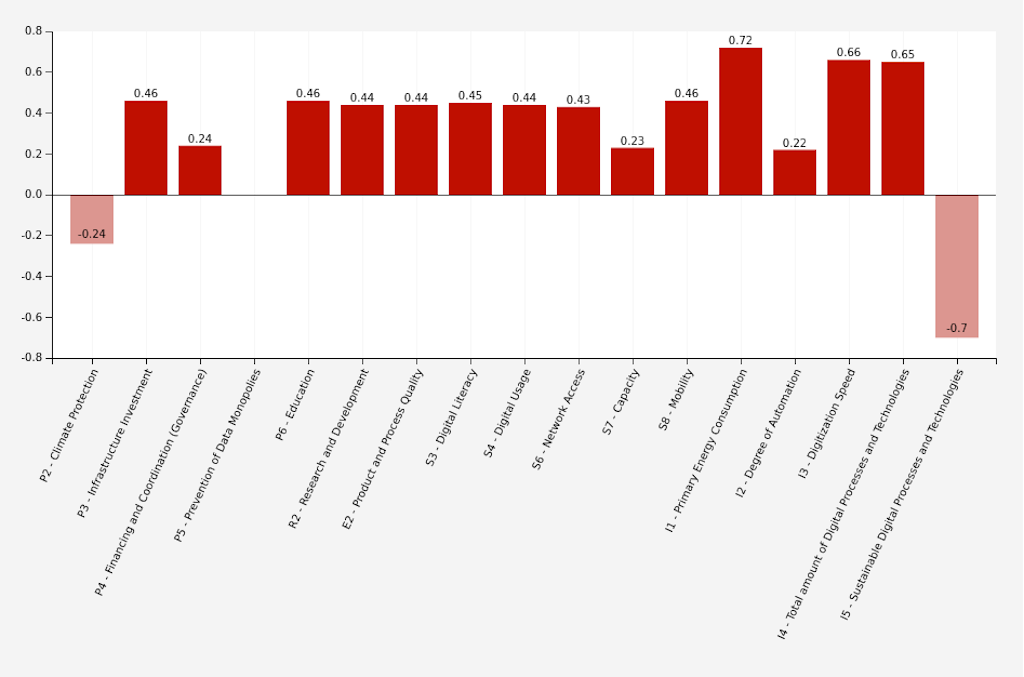}}
\caption{Simulation Results . Traditional growth-oriented scenario  (Screenshot from Mental Modeler Software)}
\label{fig4}
\end{minipage}
\end{figure*}

Figure \ref{fig3}. shows a graphical representation of the first scenario, the traditional growth-oriented scenario. It is characterized by following parameters: The state promotes infrastructure development, research and education. There is no state support for the economy, mobility is not encouraged, sustainable technologies are not promoted, and data monopolies are not avoided. There are no additional government measures for climate protection. The economy additionally supports education, infrastructure extension and the degree of automation. They invest in research for product and process quality, but do not promote sustainable technologies decisively. Research develops new technologies without promoting sustainability. The population is growing moderately and continues to pursue consumption-oriented and individual life and work models, without a significant sustainable share.
As a result, government expenses (+0.24) and expenditure on infrastructure extension (+0.46) are increasing. Data monopolies are not avoided (0). Climate protection measures are decreasing (-0.24) and the number of digital processes and technologies is increasing (+0.65), while the share of sustainable technologies is decreasing (-0.7). The digitization speed is increasing (+0.66), as well as the primary energy consumption (+0.72) and the degree of automation (+0.22). There is a higher digital literacy (+0.45), better network access (+0.43), higher capacities (+0.23) and more digital usage (+0.44) and a corresponding higher mobility (+0.46). Corresponding to an increasing product and process quality (+0.44), education (+0.46) and research (+0.44) grow. Table \ref{tab15} shows the adjacency matrix and the state vectors in scenario 1. The initial state vectors are depicted in white colour on grey ground. Figure \ref{fig4}. shows a graphical depiction of the results of scenario 1 in a bar diagram. \\

\subsection{Scenario 2 - COVID-19 Scenario (Natural Disaster)}

\begin{figure*}[htbp]
\begin{minipage}[b]{1.0\textwidth}
\centerline{\includegraphics[width=1.0\textwidth]{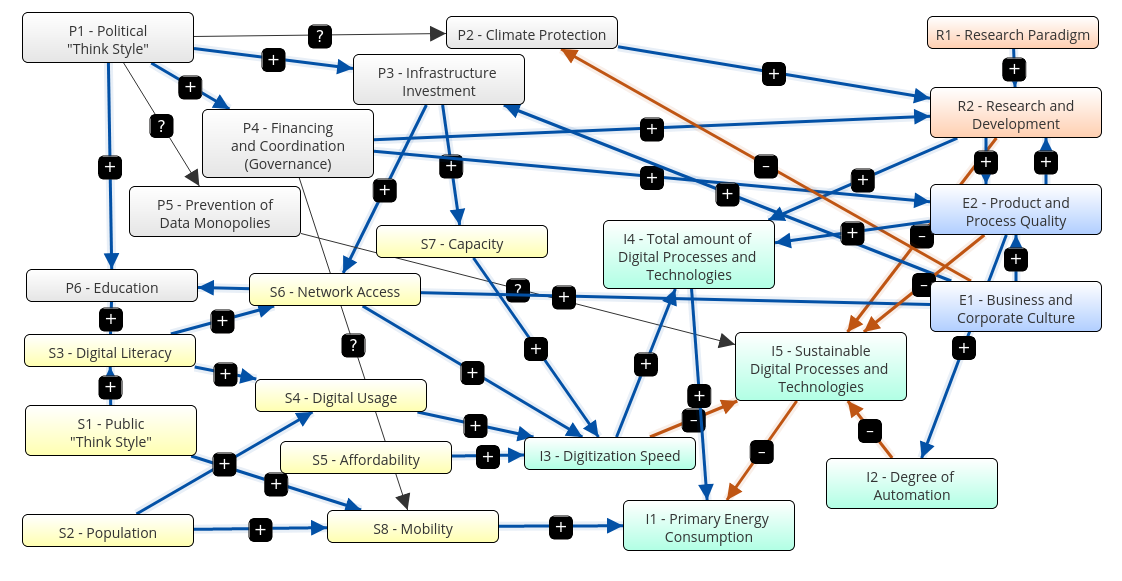}}
\caption{Graphical representation . COVID-19 scenario (Screenshot from Mental Modeler Software)}
\label{fig5}
\end{minipage}
\end{figure*}

Figure \ref{fig5}. shows a graphical representation of the second scenario, the COVID-19 scenario (natural disaster). It is characterized by following parameters: The state is promoting the extension of infrastructure and is also investing more in research. Education and digital literacy receive additional support. There is temporary state support for the economy, mobility is not promoted, sustainable technologies are not promoted, data monopolies are not avoided. There are no additional government measures for climate protection. Companies provide additional support for digital education, infrastructure extension and automation. They also invest in research for product and process quality, but without providing dedicated funding for climate protection or sustainable technologies. Research is increasingly developing new technologies without promoting sustainability. The population is growing moderately, but continues to pursue consumption-oriented and individualized models of living and working, without a significant sustainable share.

\begin{figure*}[htbp]
\begin{minipage}[b]{1.0\textwidth}
\centerline{\includegraphics[width=1.0\textwidth]{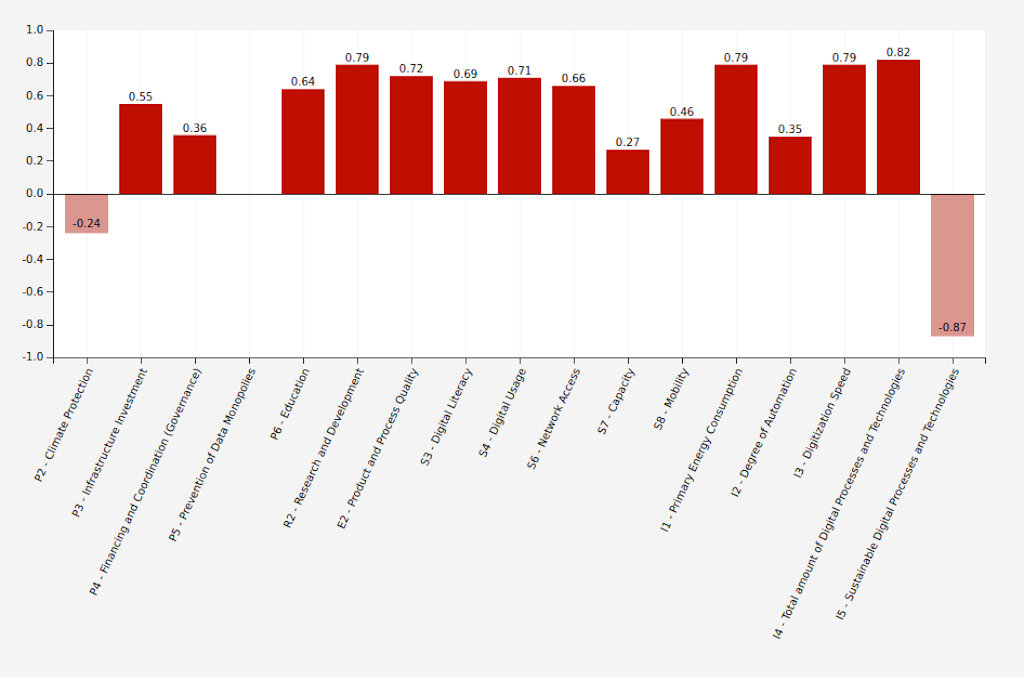}}
\caption{Simulation results . COVID-19 scenario  (Screenshot from Mental Modeler Software)}
\label{fig6}
\end{minipage}
\end{figure*}

Government expenses (+0.36) and expenditures for infrastructure extension (+0.55) are increasing. Data monopolies are not avoided (0). Climate protection measures are decreasing (-0.24). The number of digital processes and technologies is increasing (+0.82), while the share of sustainable technologies is declining (-0.87). The speed of digitization (+0.79), primary energy consumption (+0.79) and the degree of automation (+0.35) are increasing. Digital literacy is increasing (+0.69), there is better network access (+0.66), higher capacities (+0.27) and more digital usage (+0.71). Individual mobility (+0.46) and product and process quality (+0.72) increase. Education (+0.64) and research (+0.79) both grow. Table \ref{tab16} shows the adjacency matrix and the state vectors in scenario 2. The initial state vectors are depicted in white colour on grey ground. Figure \ref{fig6}. shows a graphical depiction of the results of scenario 2 in a bar diagram.\\

\subsection{Scenario 3 - Innovative and Sustainable COVID-19 Scenario}

\begin{figure*}[htbp]
\begin{minipage}[b]{1.0\textwidth}
\centerline{\includegraphics[width=1.0\textwidth]{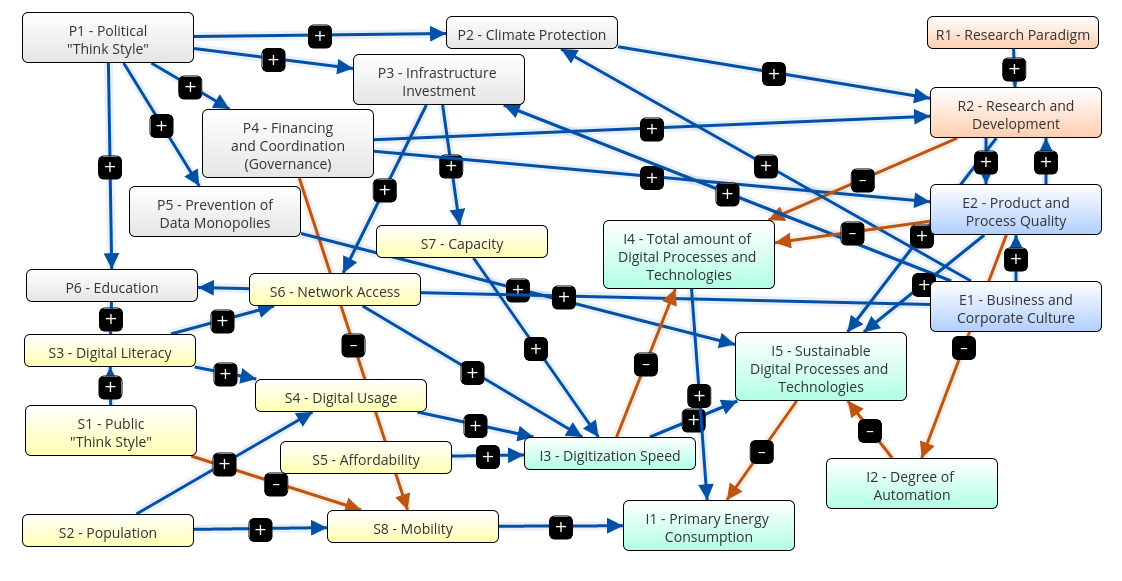}}
\caption{Graphical representation . Innovative and sustainable COVID-19 scenario (Screenshot from Mental Modeler Software)}
\label{fig7}
\end{minipage}
\end{figure*}

Figure \ref{fig7}. shows a graphical representation of the third scenario, the innovative and sustainable COVID-19 scenario. It is characterized by following parameters: There are additional government measures for climate protection. The state promotes the extension of infrastructure and avoids data monopolies. It invests in education, digital literacy and research. There is less state support for the economy. Sustainable mobility is promoted through urban structural measures (e.g. G{\"o}deritz et al. "Die gegliederte und aufgelockerte Stadt"\cite{b50}). Individual traffic is reduced. Sustainable technologies are promoted in a targeted manner. The companies support climate protection, education and infrastructure development. The degree of automation is oriented towards a socially sustainable development. The companies invest in research, product and process quality and in sustainable technologies. The population is growing moderately and pursues predominantly socially sustainable and cooperative living and working models, with a high sustainable proportion.

\begin{figure*}[htbp]
\begin{minipage}[b]{1.0\textwidth}
\centerline{\includegraphics[width=1.0\textwidth]{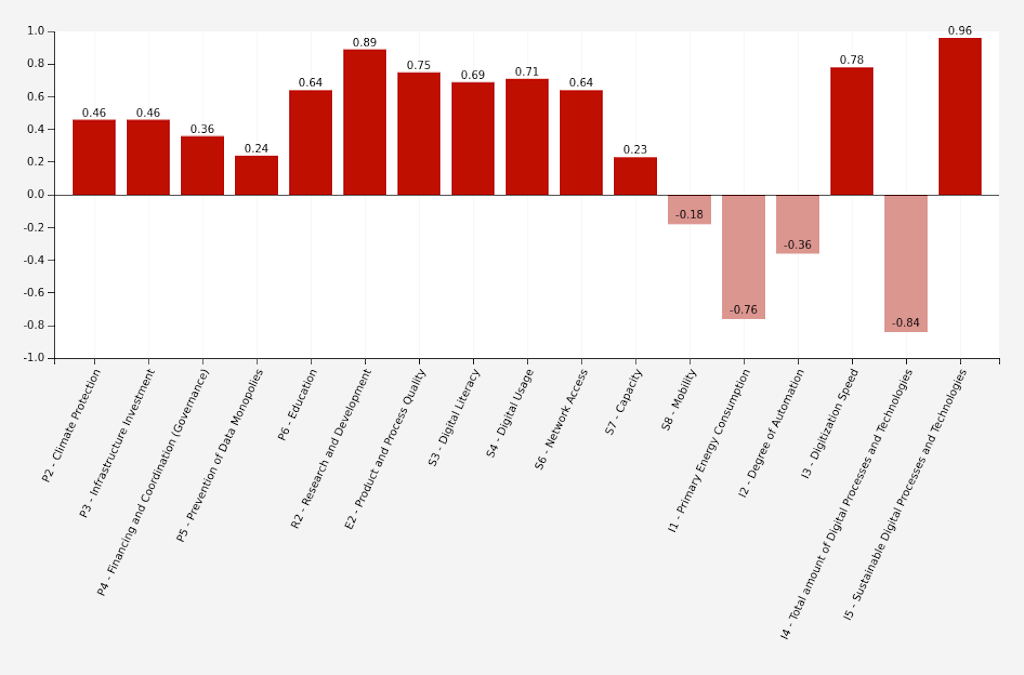}}
\caption{Simulation results . Innovative and sustainable COVID-19 scenario  (Screenshot from Mental Modeler Software)}
\label{fig8}
\end{minipage}
\end{figure*}

Government expenses (+0.36) and expenditure on infrastructure extension (+0.46) are increasing. Additional measures for climate protection (+0.46) and against data monopolies are taken (+0.24). The share of sustainable technologies (+0.96) is increasing, while the total number of digital processes and technologies is being reduced (-0.84). The digitization speed  increases (+0.78) while primary energy consumption decreases (-0.76). The degree of automation decreases (-0.36). Digital literacy is increasing (+0.69), there is better network access (+0.64), higher capacities (+0.23) and more digital usage (+0.71). Individual mobility decreases (-0.18). Product and process quality (+0.75), education (+0.64) and research (+0.89) all three grow. Table  \ref{tab17} shows the adjacency matrix and the state vectors in scenario 3. The initial state vectors are depicted in white colour on grey ground. Figure \ref{fig8}. shows a graphical depiction of the results of scenario 3 in a bar diagram. 

\newpage
\subsection{State vectors related to the different scenarios}

The different state vectors of the three alternative scenarios are shown in table  \ref{tab12}. The COVID-19 scenario 2 shows higher state vectors compared to scenario 1. These increased state vectors are applied to illustrate the higher investments, the increased expenditure in opposition to a traditional, growth-oriented scenario 1. Scenario 3 also shows these increased state vectors in many areas, but is supplemented by moderate measures for climate protection. Climate protection will be promoted, general infrastructure expenditure is reduced, data monopolies are avoided and mobility is reduced.

\begin{table}[ht]
\caption{State vectors related to the three alternative scenarios}
\begin{center}
\begin{tabularx}{1.0\columnwidth}{|p{0.01\textwidth}|p{0.09\textwidth}|p{0.09\textwidth}|p{0.095\textwidth}|X|}
\hline
\textbf{} & \textbf{} & \textbf{(1)} & \textbf{(2)} & \textbf{(3) }\\
\textbf{} & \textbf{Concept} & \textbf{Traditional Growth-oriented} & \textbf{COVID-19} & \textbf{Innovative and Sustainable COVID-19}\\
\hline
{P1} & {Political Thought Style} & {} & {} & {} \\
{P2} & {Climate Protection}  & {$\pm$0.00} & {$\pm$0.00} & {$+0.50$}\\
{P3} & {Infrastructure Investment} & {$+0.50$} & {$+0.75$} & {$+0.50$} \\
{P4} & {Financing and Coordination (Governance)}& {$+0.50$} & {$+0.75$} & {$+0.75$} \\
{P5} & {Prevention of Data Monopolies (Data Ethics)} & {$\pm$0.00} & {$\pm$0.00} & {$+0.50$} \\
{P6} & {Education} & {$+0.50$} & {$+0.75$} & {$+0.75$} \\
\hline
{R1} & {Research Paradigm} & {} & {} & {} \\
{R2} & {Research and Development} & {$+0.50$} & {$+0.75$} & {$+0.75$} \\
\hline
{E1} & {Business and Corporate Culture} & {} & {} & {} \\
{E2} & {Product and Process Quality} & {$+0.50$} & {$+0.75$} & {$+0.75$} \\
\hline
{S1} & {Public Thought Style} & {} & {} & {} \\
{S2} & {Population} & {} & {} & {} \\
{S3} & {Digital Literacy} & {$+0.50$} & {$+0.75$} & {$+0.75$} \\
{S4} & {Digital Usage} & {$+0.50$} & {$+0.75$} & {$+0.75$} \\
{S5} & {Affordability} & {} & {} & {} \\
{S6} & {Network Access}  & {} & {} & {} \\
{S7} & {Capacity} & {} & {} & {} \\
{S8} & {Mobility} & {$+0.50$} & {$+0.75$} & {$-0.50$} \\
\hline
\end{tabularx}
\label{tab12}
\end{center}
\end{table}

\subsection{Simulation results related to the different scenarios}

The results of the scenarios 1 and 2 are similar to a high degree. In both scenarios, the effort for climate protection is reduced to the same extent. The share of sustainable digital processes and technologies decreases in scenario 2 even more. Mobility remains the same, all other factors increase in scenario 2 due to the higher initial state vectors. Table  \ref{tab13} shows the results of the simulation, related to the different scenarios.

Scenario 3 differs from the first two scenarios. However, its structure is completely different. Additional measures for climate protection are taken. Conventional government expenses remain at the level of scenario 2. Expenditure on infrastructure returns to the level of scenario 1. Education remains at the same high level. Research and development grow much more strongly than in the first two scenarios. The digitization speed declines only slightly and can almost be maintained at the level of scenario 2. The same applies to network access. Digital usage remains the same. Digital literacy is increased compared to scenario 1 and corresponds to the level of scenario 2. The degree of automation is reduced. Mobility and total energy consumption will decrease. The network capacity can be maintained at the level of scenario 1. The share of sustainable digital processes and technologies can be increased significantly while the total number of digital processes and technologies decreases. Product and process quality increase.

These results speak for a reachable, innovative and research-driven, sustainable increase in efficiency in scenario 3. With almost the same effort as in scenario 2, a real social and sustainable trend reversal seems possible.

\begin{table}[htbp]
\caption{Simulation results related to the three alternative scenarios}
\begin{center}
\begin{tabularx}{1.0\columnwidth}{|p{0.01\textwidth}|p{0.09\textwidth}|p{0.09\textwidth}|p{0.095\textwidth}|X|}
\hline
\textbf{} & \textbf{} & \textbf{(1)} & \textbf{(2)} & \textbf{(3) }\\
\textbf{} & \textbf{Concept} & \textbf{Traditional Growth-oriented} & \textbf{COVID-19} & \textbf{Innovative and Sustainable COVID-19}\\
\hline
{P1} & {Political Thought Style} & {} & {} & {} \\
{P2} & {Climate Protection}  & {$-0.24$} & {$-0.24$} & {$+0.46$}\\
{P3} & {Infrastructure Investment} & {$+0.46$} & {$+0.55$} & {$+0.46$} \\
{P4} & {Financing and Coordination (Governance)}& {$+0.24$} & {$+0.36$} & {$+0.36$} \\
{P5} & {Prevention of Data Monopolies (Data Ethics)} & {$\pm$0.00} & {$\pm$0.00} & {$+0.24$} \\
{P6} & {Education} & {$+0.46$} & {$+0.64$} & {$+0.64$} \\
\hline
{R1} & {Research Paradigm} & {} & {} & {} \\
{R2} & {Research and Development} & {$+0.44$} & {$+0.79$} & {$+0.89$} \\
\hline
{E1} & {Business and Corporate Culture} & {} & {} & {} \\
{E2} & {Product and Process Quality} & {$+0.44$} & {$+0.72$} & {$+0.75$} \\
\hline
{S1} & {Public Thought Style} & {} & {} & {} \\
{S2} & {Population} & {} & {} & {} \\
{S3} & {Digital Literacy} & {$+0.45$} & {$+0.69$} & {$+0.69$} \\
{S4} & {Digital Usage} & {$+0.44$} & {$+0.71$} & {$+0.71$} \\
{S5} & {Affordability} & {} & {} & {} \\
{S6} & {Network Access} & {$+0.43$} & {$+0.66$} & {$+0.64$} \\
{S7} & {Capacity} & {$+0.23$} & {$+0.27$} & {$+0.23$} \\
{S8} & {Mobility} & {$+0.46$} & {$+0.46$} & {$-0.18$} \\
\hline
{I1} & {Primary Energy Consumption} & {$+0.72$} & {$+0.79$} & {$+0.76$} \\
{I2} & {Degree of Automation} & {$+0.22$} & {$+0.35$} & {$+0.36$} \\
{I3} & {Digitization Speed} & {$+0.66$} & {$+0.79$} & {$+0.78$} \\
{I4} & {Total amount of digital processes and technologies} & {$+0.65$} & {$+0.82$} & {$+0.84$} \\
{I5} & {Sustainable digital processes and technologies} & {$-0.70$} & {$-0.87$} & {$+0.94$} \\
\hline
\end{tabularx}
\label{tab13}
\end{center}
\end{table}

\subsection{Network parameters and centrality}

\begin{figure*}[htbp]
\begin{minipage}[b]{1.0\textwidth}
\centerline{\includegraphics[width=1.0\textwidth]{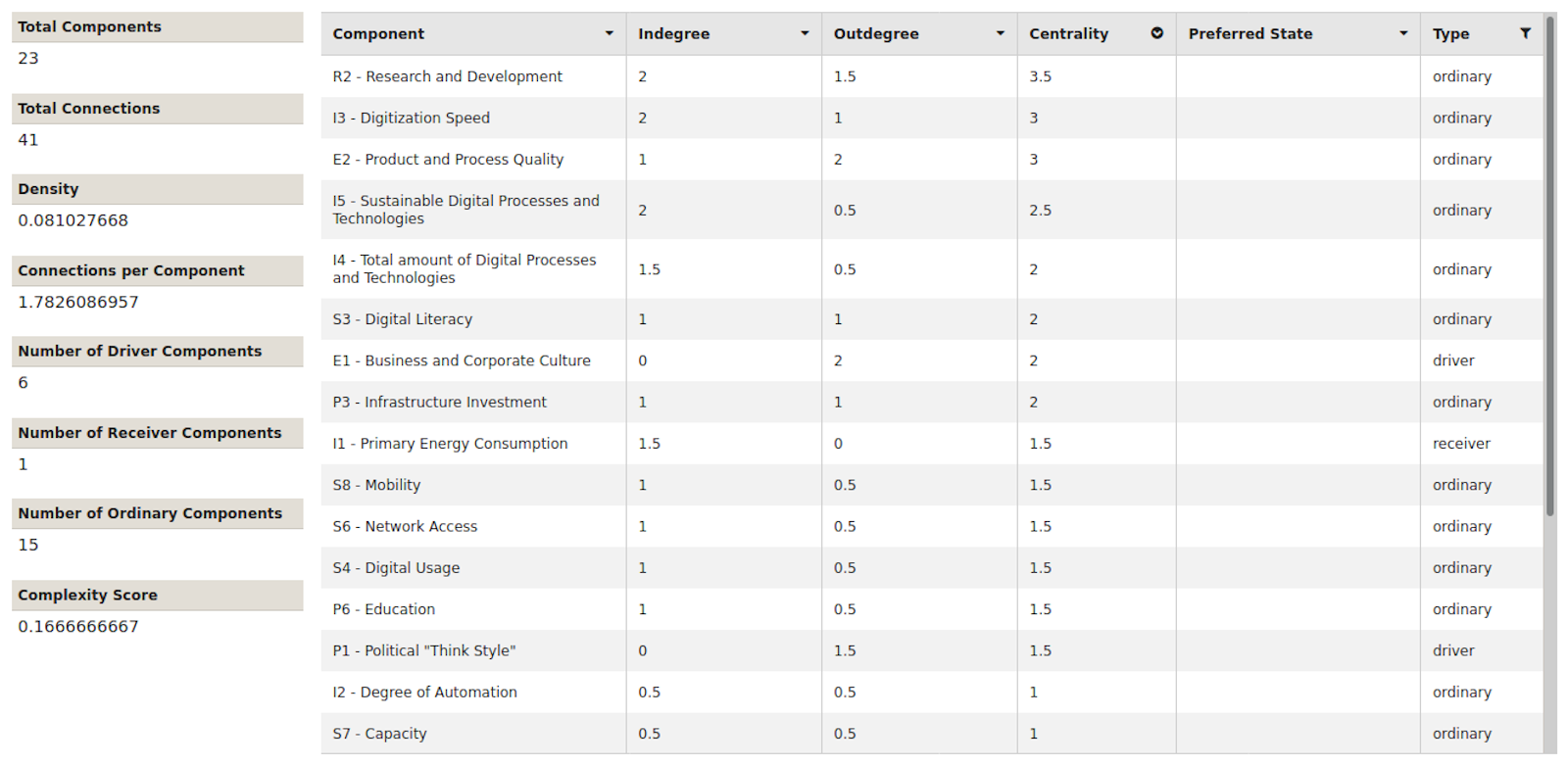}}
\caption{Traditional growth-oriented scenario . Network parameters and centrality (Screenshot from Mental Modeler Software)}
\label{fig9}
\end{minipage}
\end{figure*}

\begin{figure*}[htbp]
\begin{minipage}[b]{1.0\textwidth}
\centerline{\includegraphics[width=1.0\textwidth]{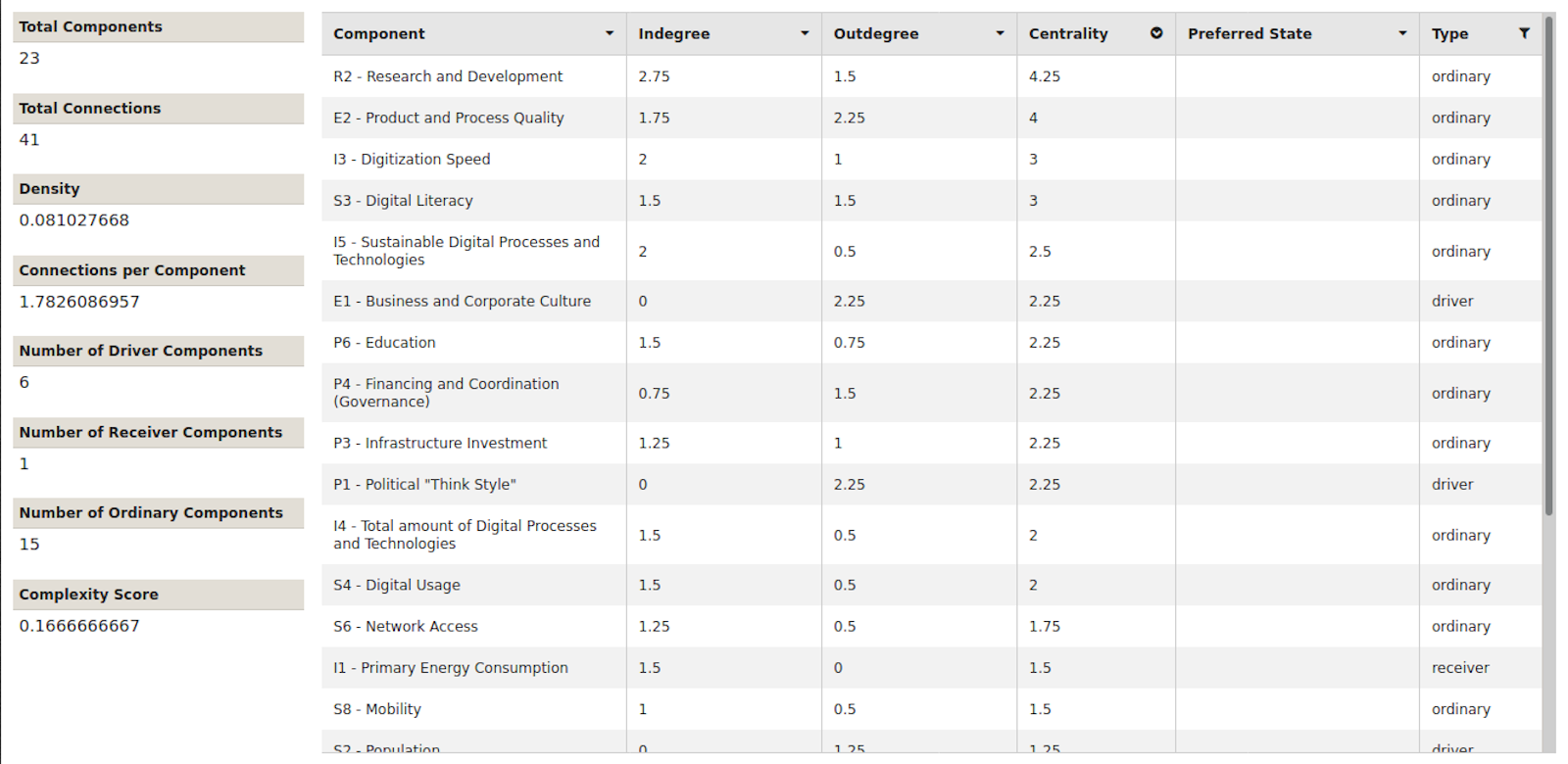}}
\caption{COVID-19 scenario (natural disaster) . Network parameters and centrality (Screenshot from Mental Modeler Software)}
\label{fig10}
\end{minipage}
\end{figure*}

\begin{figure*}[htbp]
\begin{minipage}[b]{1.0\textwidth}
\centerline{\includegraphics[width=1.0\textwidth]{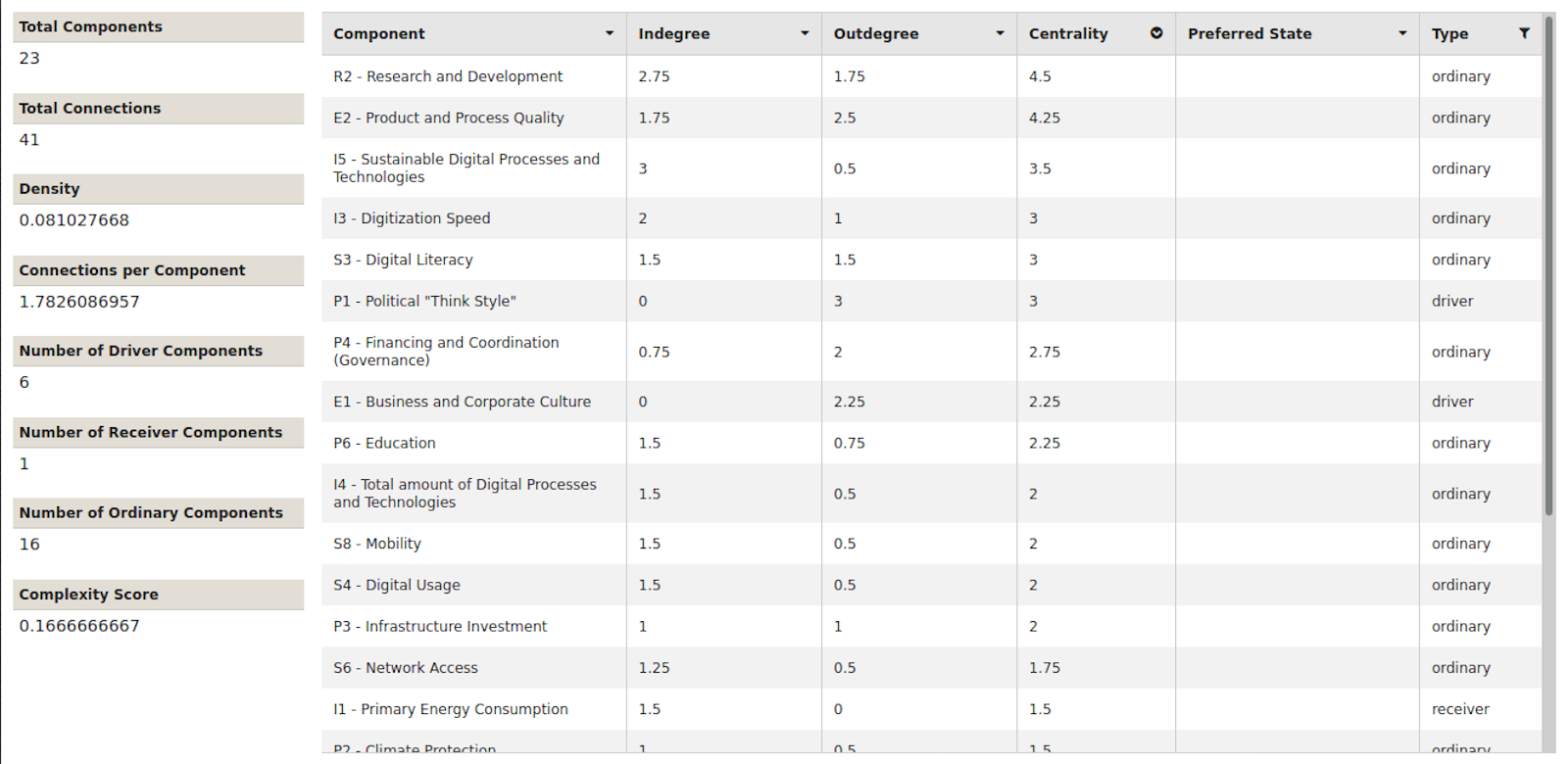}}
\caption{Innovative and sustainable COVID-19 scenario . Network parameters and centrality (Screenshot from Mental Modeler Software)}
\label{fig11}
\end{minipage}
\end{figure*}

The network parameters in all three scenarios are completely identical, as shown in figures \ref{fig9}., \ref{fig10}. and \ref{fig11}. All figures and tables shown in this paper you can easily access in the original version through the provided XML-files in the supplementary materials. They all have the same number of components and connections. They have the same density and the same number of connections per component. All three models have exactly the same complexity score and the same six driver components. Sorting the nodes according to their centrality, you can see the following, as shown in table \ref{tab14}: All three scenarios differ from each other in the order and evaluation of the centrality parameters.

R2 (Research and Development) has the highest centrality rating in all three scenarios. The centrality of research and development also increases, in scenarios 1 - 3. In scenario 1 (traditional, growth-oriented scenario), the digitization speed is in second place, on a par with product and process quality. Both followed by the share of sustainable processes and technologies. The total number of digital processes and technologies moves up to the fifth place.

Interestingly, in scenario 2 (COVID-19 scenario), product and process quality moves up to second place. Third place is shared by the digitization speed and digital literacy. The order has clearly changed compared to the first scenario. The share of sustainable processes and technologies is in fifth place.

In scenario 3 (innovative and sustainable COVID-19 scenario), product and process quality retains second place. However, the third place is now clearly followed by the share of sustainable processes and technologies. The digitization speed and digital literacy together fall back to the fourth place.

\begin{table}[htbp]
\caption{Ranking of concepts due to network centrality}
\begin{center}
\begin{tabularx}{1.0\columnwidth} {|p{0.05\textwidth}|X|p{0.05\textwidth}|X|p{0.05\textwidth}|X|}
\hline
\textbf{(1)} & \textbf{} & \textbf{(2)} & \textbf{} & \textbf{(3)} & \textbf{}\\
\textbf{Trad. Growth-oriented} & \textbf{Concept} & \textbf{COVID-19} & \textbf{Concept} & \textbf{Innov. and Sustainable COVID-19} & \textbf{Concept} \\
\hline
{$3.50$} & {R2} & {$4.25$} & {R2} & {$4.50$}  & {R2}\\
{} & {Research and Development} & {} & {Research and Development} & {}  & {Research and Development}\\
{} & {} & {} & {} & {}  & {}\\
{$3.00$} & {I3} & {$4.00$} & {E2} & {$4.25$}  & {E2}\\
{} & {Digitization Speed} & {} & {Product and Process Quality} & {}  & {Product and Process Quality}\\
{} & {} & {} & {} & {}  & {}\\
{$3.00$} & {E2} & {$3.00$} & {I3} & {$3.50$}  & {I5}\\
{} & {Product and Process Quality} & {} & {Digitization Speed} & {}  & {Sustainable digital processes and technologies}\\
{} & {} & {} & {} & {}  & {}\\
{$2.50$} & {I5} & {$3.00$} & {S3} & {$3.00$}  & {I3}\\
{} & {Sustainable digital processes and technologies} & {} & {Digital Literacy} & {}  & {Digitization Speed}\\
{} & {} & {} & {} & {}  & {}\\
{} & {} & {} & {} & {}  & {}\\
{$2.00$} & {I4} & {$2.50$} & {I5} & {$3.00$}  & {S3}\\
{} & {Total amount of digital processes and technologies} & {} & {Sustainable digital processes and technologies} & {}  & {Digital Literacy}\\
{} & {} & {} & {} & {}  & {}\\
\hline
\end{tabularx}
\label{tab14}
\end{center}
\end{table}

\newpage
\section{Discussion and Conclusions}
According to the results of this study, a sustainable and social turnaround is possible. It is described how a reduction of primary energy consumption and a simultaneous increase in the share of sustainable digital processes and technologies can be achieved with the same or a slightly higher effort.
The order of the concepts according to network centrality shows much of what we can already observe in reality today. In all three scenarios, research drives the development of new technologies and is at the center of the digital transformation. All three scenarios have different initial state vectors and digitization speeds. In the traditional and growth-oriented scenario 1, the focus is on digitization speed and on the quality of economic processes and products. The share of sustainable processes and technologies comes after these two factors. This scenario best reflects the well-known and widespread efficiency-oriented and fordist production factors of performance, quality and acceleration. In COVID-19 scenario 2, the quality of products and processes takes second place, followed by digitization speed. With higher initial state vectors and a higher effort the product and process quality increases. The digitization speed still plays an important role, but loses centrality. It is shown that there is a factual connection to a general "slowdown" in the course of the current pandemic. Interestingly, digital literacy now appears to be more prominent. The share of sustainable processes is slipping further behind. The simulation results show, that in times of COVID-19, sustainability is not necessarily written in capital letters, and we can observe this as well in reality. In scenario 3, the innovative and sustainable COVID-19 scenario, research is also at the top of the list, followed by product and process quality. However, the share of sustainable processes and technologies now takes on a central position. The digitization speed and digital literacy are moving further back, but are still in the focus of our interests. Taking into account digital education and a moderate digitization speed, further improvement of our products and processes can lead to sustainable quality and real innovation?
For architecture an urban planning these results are very important. Innovative digital transformation will only be possible by a general, integrative and coordinated change of "thought styles" of all stakeholders involved. Innovative sustainability can be reached through more intensive and target-oriented research and development from both sides, economy as well as science. Research and development will be the basis for really innovative and sustainable products and services. Therefore we will need a reliable classification of sustainable technologies. Additionally an adequate digital literacy and responsible public handling of resources will be needed. Mobility can be optimized through appropriate urban structure planning. It will not only be the question of a new category of vehicles. We all will have to review our lifestyle. Urban planning will have to intensively review the city structures for innovative and sustainable principles and generate ideas for sustainable city quarters of the future. Where do we want to live, where do we work, do we really have to travel that much? And we all will see processes slowing down to a socially compatible, sustainable and human level.
Technological development already fulfils many future scenarios today. Even if we are not far away from the adaptivity and self-organisation of digital systems, there is still a long way to go for automatic context perception and comprehensive standards. It would also be desirable to have a solid, worldwide non-profit organization in order to make high-quality data openly available. For weather data, the World Meteorological Organization of the UN has been in existence for 70 years. Currently, there are also few standardized assessments of digital processes and technologies. Often no clear distinction is made between conventional and sustainable digital technologies. Often the mutual basics are missing. A catalog of innovative and sustainable social processes and technologies for targeted promotion and financing would be desirable. The degree of automation and social compatibility of innovative processes and technologies should also play a key role in this context. A central organization at global or european level could certainly make a significant contribution to this.
For future work there are many interesting perspectives. The ETSIE model provided a first proof of concept comparing three scenarios of digital transformation using a relatively simple software basis. Of course the figures can be improved to illustrate the advantages of the innovative and sustainable model. In the next steps the number of scenarios could be extended using another software platform or a combination of advanced software to reach the whole realm of the experimental simulation's testbed. Generating more scenarios. using learning algorithms and evolutionary optimization and evaluation will lead to refinement and general improvement of the model and should provide new and more detailed information on the general process of digital transformation in architecture andurban planning.
So, will COVID-19 accelerate digitization? To a certain extent yes. This study has shown that COVID-19 can be an accelerator of digitization, but is no guarantee for sustainable and high quality innovation. Similar to a natural disaster, COVID-19 can be presented as just one of many possible digitization scenarios. The investments and all the stakeholders decision will show which direction we want to take. Whether we want to continue to pursue primarily commercial interests, or whether we see the sustainable benefits of a solidary community in a healthy and functioning, really innovative environment. It will not be enough to digitize the existing growth-driven economy. A fundamental and innovative change is necessary to avert a dramatic climate catastrophe and to solve our social problems. This change must be supported by all stakeholders in the same way and can be implemented with nearly the same or a moderate higher effort.
The current crisis is an opportunity for new ideas, for the meaningful restructuring of creative and productive processes. But this does not only include investment in infrastructure and digital literacy. How we want to keep it, that way is up to us to decide. The question remains - in which world do we want to live?

\section*{Acknowledgment}

The author would like to thank Theresa Ramisch and Alexander Gutzmer from Baumeister Magazine at Georg Media GmbH. in Munich, who asked me to have a closer look at this interesting topic. My thanks also go to my partner Cornelia for her patience and support developing this small research project and her valuable advice. Last but not least, my sincere gratitude goes to Gudrun Klinker at TUM, who kindly made this publication possible.

\section{Appendix}


\begin{table*}[htbp]
\begin{minipage}[b]{1.0\textwidth}
\caption{Adjacency matrix and state vectors in scenario 1 (Traditional growth-oriented scenario)}
\begin{center}
\begin{tabularx}{1.0\textwidth}{|X|X|X|X|X|X|X|X|X|X|X|X|X|X|X|X|X|X|X|X|X|X|X|X|}
\hline
{} & {P1} & {P2} & {P3} & {P4} & {P5} & {P6} & {R1} & {R2} & {E1} & {E2} & {S1} & {S2} & {S3} & {S4} & {S5} & {S6} & {S7} & {S8} & {I1} & {I2} & {I3} & {I4} & {I5}\\
\hline
{P1} & {} & {0} & {0.5} & {0.5} & {0} & {0.5} & {} & {} & {} & {} & {} & {} & {} & {} & {} & {} & {} & {} & {} & {} & {} & {} & {}\\
\hline
{P2} & {} & {} & {} & {} & {} & {} & {} & {0.5} & {} & {} & {} & {} & {} & {} & {} & {} & {} & {} & {} & {} & {} & {} & {}\\
\hline
{P3} & {} & {} & {} & {} & {} & {} & {} & {} & {} & {} & {} & {} & {} & {} & {} & {0.5} & {0.5} & {} & {} & {} & {} & {} & {}\\
\hline
{P4} & {} & {} & {} & {} & {} & {} & {} & {0.5} & {} & {0} & {} & {} & {} & {} & {} & {} & {} & {0} & {} & {} & {} & {} & {}\\
\hline
{P5} & {} & {} & {} & {} & {} & {} & {} & {} & {} & {} & {} & {} & {} & {} & {} & {} & {} & {} & {} & {} & {} & {} & {0}\\
\hline
{P6} & {} & {} & {} & {} & {} & {} & {} & {} & {} & {} & {} & {} & {0.5} & {} & {} & {} & {} & {} & {} & {} & {} & {} & {}\\
\hline
{R1} & {} & {} & {} & {} & {} & {} & {} & {0.5} & {} & {} & {} & {} & {} & {} & {} & {} & {} & {} & {} & {} & {} & {} & {}\\
\hline
{R2} & {} & {} & {} & {} & {} & {} & {} & {} & {} & {0.5} & {} & {} & {} & {} & {} & {} & {} & {} & {} & {} & {} & {0.5} & {$-0.5$}\\
\hline
{E1} & {} & {$-0.5$} & {0.5} & {} & {} & {0.5} & {} & {} & {} & {0.5} & {} & {} & {} & {} & {} & {} & {} & {} & {} & {} & {} & {} & {}\\
\hline
{E2} & {} & {} & {} & {} & {} & {} & {} & {0.5} & {} & {} & {} & {} & {} & {} & {} & {} & {} & {} & {} & {0.5} & {} & {0.5} & {$-0.5$}\\
\hline
{S1} & {} & {} & {} & {} & {} & {} & {} & {} & {} & {} & {} & {} & {0.5} & {} & {} & {} & {} & {0.5} & {} & {} & {} & {} & {}\\
\hline
{S2} & {} & {} & {} & {} & {} & {} & {} & {} & {} & {} & {} & {} & {} & {0.5} & {} & {} & {} & {0.5} & {} & {} & {} & {} & {}\\
\hline
{S3} & {} & {} & {} & {} & {} & {} & {} & {} & {} & {} & {} & {} & {} & {0.5} & {} & {0.5} & {} & {} & {} & {} & {} & {} & {}\\
\hline
{S4} & {} & {} & {} & {} & {} & {} & {} & {} & {} & {} & {} & {} & {} & {} & {} & {} & {} & {} & {} & {} & {0.5} & {} & {}\\
\hline
{S5} & {} & {} & {} & {} & {} & {} & {} & {} & {} & {} & {} & {} & {} & {} & {} & {} & {} & {} & {} & {} & {0.5} & {} & {}\\
\hline
{S6} & {} & {} & {} & {} & {} & {} & {} & {} & {} & {} & {} & {} & {} & {} & {} & {} & {} & {} & {} & {} & {0.5} & {} & {}\\
\hline
{S7} & {} & {} & {} & {} & {} & {} & {} & {} & {} & {} & {} & {} & {} & {} & {} & {} & {} & {} & {} & {} & {0.5} & {} & {}\\
\hline
{S8} & {} & {} & {} & {} & {} & {} & {} & {} & {} & {} & {} & {} & {} & {} & {} & {} & {} & {} & {0.5} & {} & {} & {} & {}\\
\hline
{I1} & {} & {} & {} & {} & {} & {} & {} & {} & {} & {} & {} & {} & {} & {} & {} & {} & {} & {} & {} & {} & {} & {} & {}\\
\hline
{I2} & {} & {} & {} & {} & {} & {} & {} & {} & {} & {} & {} & {} & {} & {} & {} & {} & {} & {} & {} & {} & {} & {} & {$-0.5$}\\
\hline
{I3} & {} & {} & {} & {} & {} & {} & {} & {} & {} & {} & {} & {} & {} & {} & {} & {} & {} & {} & {} & {} & {} & {0.5} & {$-0.5$}\\
\hline
{I4} & {} & {} & {} & {} & {} & {} & {} & {} & {} & {} & {} & {} & {} & {} & {} & {} & {} & {} & {0.5} & {} & {} & {} & {}\\
\hline
{I5} & {} & {} & {} & {} & {} & {} & {} & {} & {} & {} & {} & {} & {} & {} & {} & {} & {} & {} & {$-0.5$} & {} & {} & {} & {}\\
\hline
\end{tabularx}
\label{tab15}
\end{center}
\end{minipage}
\end{table*}


\begin{table*}[htbp]
\begin{minipage}[b]{1.0\textwidth}
\caption{Adjacency matrix and state vectors in scenario 2 (COVID-19 scenario or natural disaster)}
\begin{center}
\begin{tabularx}{1.0\textwidth}{|X|X|X|X|X|X|X|X|X|X|X|X|X|X|X|X|X|X|X|X|X|X|X|X|}
\hline
{} & {P1} & {P2} & {P3} & {P4} & {P5} & {P6} & {R1} & {R2} & {E1} & {E2} & {S1} & {S2} & {S3} & {S4} & {S5} & {S6} & {S7} & {S8} & {I1} & {I2} & {I3} & {I4} & {I5}\\
\hline
{P1} & {} & {0} & {0.75} & {0.75} & {0} & {0.75} & {} & {} & {} & {} & {} & {} & {} & {} & {} & {} & {} & {} & {} & {} & {} & {} & {}\\
\hline
{P2} & {} & {} & {} & {} & {} & {} & {} & {0.5} & {} & {} & {} & {} & {} & {} & {} & {} & {} & {} & {} & {} & {} & {} & {}\\
\hline
{P3} & {} & {} & {} & {} & {} & {} & {} & {} & {} & {} & {} & {} & {} & {} & {} & {0.5} & {0.5} & {} & {} & {} & {} & {} & {}\\
\hline
{P4} & {} & {} & {} & {} & {} & {} & {} & {0.75} & {} & {0.75} & {} & {} & {} & {} & {} & {} & {} & {0} & {} & {} & {} & {} & {}\\
\hline
{P5} & {} & {} & {} & {} & {} & {} & {} & {} & {} & {} & {} & {} & {} & {} & {} & {} & {} & {} & {} & {} & {} & {} & {0}\\
\hline
{P6} & {} & {} & {} & {} & {} & {} & {} & {} & {} & {} & {} & {} & {0.75} & {} & {} & {} & {} & {} & {} & {} & {} & {} & {}\\
\hline
{R1} & {} & {} & {} & {} & {} & {} & {} & {0.75} & {} & {} & {} & {} & {} & {} & {} & {} & {} & {} & {} & {} & {} & {} & {}\\
\hline
{R2} & {} & {} & {} & {} & {} & {} & {} & {} & {} & {0.5} & {} & {} & {} & {} & {} & {} & {} & {} & {} & {} & {} & {0.5} & {$-0.5$}\\
\hline
{E1} & {} & {$-0.5$} & {0.5} & {} & {} & {0.75} & {} & {} & {} & {0.5} & {} & {} & {} & {} & {} & {} & {} & {} & {} & {} & {} & {} & {}\\
\hline
{E2} & {} & {} & {} & {} & {} & {} & {} & {0.75} & {} & {} & {} & {} & {} & {} & {} & {} & {} & {} & {} & {0.5} & {} & {0.5} & {$-0.5$}\\
\hline
{S1} & {} & {} & {} & {} & {} & {} & {} & {} & {} & {} & {} & {} & {0.75} & {} & {} & {} & {} & {0.75} & {} & {} & {} & {} & {}\\
\hline
{S2} & {} & {} & {} & {} & {} & {} & {} & {} & {} & {} & {} & {} & {} & {0.75} & {} & {} & {} & {0.5} & {} & {} & {} & {} & {}\\
\hline
{S3} & {} & {} & {} & {} & {} & {} & {} & {} & {} & {} & {} & {} & {} & {0.75} & {} & {0.75} & {} & {} & {} & {} & {} & {} & {}\\
\hline
{S4} & {} & {} & {} & {} & {} & {} & {} & {} & {} & {} & {} & {} & {} & {} & {} & {} & {} & {} & {} & {} & {0.5} & {} & {}\\
\hline
{S5} & {} & {} & {} & {} & {} & {} & {} & {} & {} & {} & {} & {} & {} & {} & {} & {} & {} & {} & {} & {} & {0.5} & {} & {}\\
\hline
{S6} & {} & {} & {} & {} & {} & {} & {} & {} & {} & {} & {} & {} & {} & {} & {} & {} & {} & {} & {} & {} & {0.5} & {} & {}\\
\hline
{S7} & {} & {} & {} & {} & {} & {} & {} & {} & {} & {} & {} & {} & {} & {} & {} & {} & {} & {} & {} & {} & {0.5} & {} & {}\\
\hline
{S8} & {} & {} & {} & {} & {} & {} & {} & {} & {} & {} & {} & {} & {} & {} & {} & {} & {} & {} & {0.5} & {} & {} & {} & {}\\
\hline
{I1} & {} & {} & {} & {} & {} & {} & {} & {} & {} & {} & {} & {} & {} & {} & {} & {} & {} & {} & {} & {} & {} & {} & {}\\
\hline
{I2} & {} & {} & {} & {} & {} & {} & {} & {} & {} & {} & {} & {} & {} & {} & {} & {} & {} & {} & {} & {} & {} & {} & {$-0.5$}\\
\hline
{I3} & {} & {} & {} & {} & {} & {} & {} & {} & {} & {} & {} & {} & {} & {} & {} & {} & {} & {} & {} & {} & {} & {0.5} & {$-0.5$}\\
\hline
{I4} & {} & {} & {} & {} & {} & {} & {} & {} & {} & {} & {} & {} & {} & {} & {} & {} & {} & {} & {0.5} & {} & {} & {} & {}\\
\hline
{I5} & {} & {} & {} & {} & {} & {} & {} & {} & {} & {} & {} & {} & {} & {} & {} & {} & {} & {} & {$-0.5$} & {} & {} & {} & {}\\
\hline
\end{tabularx}
\label{tab16}
\end{center}
\end{minipage}
\end{table*}

\begin{table*}[htbp]
\begin{minipage}[b]{1.0\textwidth}
\caption{Adjacency matrix and state vectors in scenario 3 (Innovative and sustainable COVID-19 scenario)}
\begin{center}
\begin{tabularx}{1.0\textwidth}{|X|X|X|X|X|X|X|X|X|X|X|X|X|X|X|X|X|X|X|X|X|X|X|X|}
\hline
{} & {P1} & {P2} & {P3} & {P4} & {P5} & {P6} & {R1} & {R2} & {E1} & {E2} & {S1} & {S2} & {S3} & {S4} & {S5} & {S6} & {S7} & {S8} & {I1} & {I2} & {I3} & {I4} & {I5}\\
\hline
{P1} & {} & {0.5} & {0.5} & {0.75} & {0.5} & {0.75} & {} & {} & {} & {} & {} & {} & {} & {} & {} & {} & {} & {} & {} & {} & {} & {} & {}\\
\hline
{P2} & {} & {} & {} & {} & {} & {} & {} & {0.5} & {} & {} & {} & {} & {} & {} & {} & {} & {} & {} & {} & {} & {} & {} & {}\\
\hline
{P3} & {} & {} & {} & {} & {} & {} & {} & {} & {} & {} & {} & {} & {} & {} & {} & {0.5} & {0.5} & {} & {} & {} & {} & {} & {}\\
\hline
{P4} & {} & {} & {} & {} & {} & {} & {} & {0.75} & {} & {0.75} & {} & {} & {} & {} & {} & {} & {} & {$-0.5$} & {} & {} & {} & {} & {}\\
\hline
{P5} & {} & {} & {} & {} & {} & {} & {} & {} & {} & {} & {} & {} & {} & {} & {} & {} & {} & {} & {} & {} & {} & {} & {0.5}\\
\hline
{P6} & {} & {} & {} & {} & {} & {} & {} & {} & {} & {} & {} & {} & {0.75} & {} & {} & {} & {} & {} & {} & {} & {} & {} & {}\\
\hline
{R1} & {} & {} & {} & {} & {} & {} & {} & {0.75} & {} & {} & {} & {} & {} & {} & {} & {} & {} & {} & {} & {} & {} & {} & {}\\
\hline
{R2} & {} & {} & {} & {} & {} & {} & {} & {} & {} & {0.5} & {} & {} & {} & {} & {} & {} & {} & {} & {} & {} & {} & {$-0.5$} & {0.75}\\
\hline
{E1} & {} & {0.5} & {0.5} & {} & {} & {0.75} & {} & {} & {} & {0.5} & {} & {} & {} & {} & {} & {} & {} & {} & {} & {} & {} & {} & {}\\
\hline
{E2} & {} & {} & {} & {} & {} & {} & {} & {0.75} & {} & {} & {} & {} & {} & {} & {} & {} & {} & {} & {} & {$-0.5$} & {} & {$-0.5$} & {0.75}\\
\hline
{S1} & {} & {} & {} & {} & {} & {} & {} & {} & {} & {} & {} & {} & {0.75} & {} & {} & {} & {} & {$-0.5$} & {} & {} & {} & {} & {}\\
\hline
{S2} & {} & {} & {} & {} & {} & {} & {} & {} & {} & {} & {} & {} & {} & {0.75} & {} & {} & {} & {0.5} & {} & {} & {} & {} & {}\\
\hline
{S3} & {} & {} & {} & {} & {} & {} & {} & {} & {} & {} & {} & {} & {} & {0.75} & {} & {0.75} & {} & {} & {} & {} & {} & {} & {}\\
\hline
{S4} & {} & {} & {} & {} & {} & {} & {} & {} & {} & {} & {} & {} & {} & {} & {} & {} & {} & {} & {} & {} & {0.5} & {} & {}\\
\hline
{S5} & {} & {} & {} & {} & {} & {} & {} & {} & {} & {} & {} & {} & {} & {} & {} & {} & {} & {} & {} & {} & {0.5} & {} & {}\\
\hline
{S6} & {} & {} & {} & {} & {} & {} & {} & {} & {} & {} & {} & {} & {} & {} & {} & {} & {} & {} & {} & {} & {0.5} & {} & {}\\
\hline
{S7} & {} & {} & {} & {} & {} & {} & {} & {} & {} & {} & {} & {} & {} & {} & {} & {} & {} & {} & {} & {} & {0.5} & {} & {}\\
\hline
{S8} & {} & {} & {} & {} & {} & {} & {} & {} & {} & {} & {} & {} & {} & {} & {} & {} & {} & {} & {0.5} & {} & {} & {} & {}\\
\hline
{I1} & {} & {} & {} & {} & {} & {} & {} & {} & {} & {} & {} & {} & {} & {} & {} & {} & {} & {} & {} & {} & {} & {} & {}\\
\hline
{I2} & {} & {} & {} & {} & {} & {} & {} & {} & {} & {} & {} & {} & {} & {} & {} & {} & {} & {} & {} & {} & {} & {} & {$-0.5$}\\
\hline
{I3} & {} & {} & {} & {} & {} & {} & {} & {} & {} & {} & {} & {} & {} & {} & {} & {} & {} & {} & {} & {} & {} & {$-0.5$} & {0.5}\\
\hline
{I4} & {} & {} & {} & {} & {} & {} & {} & {} & {} & {} & {} & {} & {} & {} & {} & {} & {} & {} & {0.5} & {} & {} & {} & {}\\
\hline
{I5} & {} & {} & {} & {} & {} & {} & {} & {} & {} & {} & {} & {} & {} & {} & {} & {} & {} & {} & {$-0.5$} & {} & {} & {} & {}\\
\hline
\end{tabularx}
\label{tab17}
\end{center}
\end{minipage}
\end{table*}

\newpage

\begin{figure}[htbp]
\includegraphics[width=0.3\columnwidth]{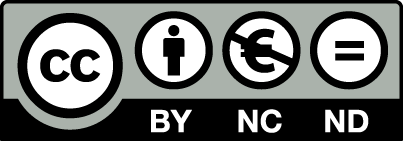}
\end{figure}

© 2021 by the author. This publication is licensed under the terms and conditions of the Creative Commons Attribution (CC BY-NC-ND) license (https://creativecommons.org/licenses/by-nc-nd/4.0/).



\begin{thebibliography}{00}
\bibitem{b1} Coronavirus COVID-19 Global Cases by the Center for Systems Science and Engineering (CSSE) at Johns Hopkins University (JHU) in Baltimore, Maryland / USA.
\bibitem{b2} "Die Welt und die Gesellschaft in Deutschland wird sich verändern", Deutschlandfunk 2020.
\bibitem{b3} KMK-Präsidentin erwartet Schub für Digitalisierung an Schulen, Deutschlandfunk 2020.
\bibitem{b4} Schmidt, P.; Ebner, S. Petition gegen den Fernunterricht, Münchner Merkur 2020 Nr. 71, 25-03-2020, p. 16.
\bibitem{b5} Zadeh L. A., The roles of fuzzy logic and soft computing in the conception, design, and deployment of intelligence systems, in: Nwana H. S., Azarmi N. (eds) 1997. Software agents and soft computing: towards enhancing machine intelligence concepts and applications. Lecture notes in computer science, vol. 1198, pp 83–90.
\bibitem{b6} Zadeh L. A., What is Soft Computing. 1997 Soft Comput 1:1–2.
\bibitem{b7} Bonissone P., Soft computing: the convergence of emerging reasoning technologies. 1997 Soft Comput 1:6–18.
\bibitem{b8} Jain L., Soft computing techniques in knowledge-based intelligent engineering systems: approaches and applications. 1997 Studies in fuzziness and soft computing, vol. 10, Springer, Berlin Heidelberg New York.
\bibitem{b9} Papageorgiou, E.I., Salmeron, J.L.: A review of fuzzy cognitive maps research during the last decade. 2013 IEEE Trans. Fuzzy Syst. 21(1), 69–79.
\bibitem{b10}Özesmi, U., Özesmi, S.: A participatory approach to ecosystem conservation: fuzzy cognitive maps and stakeholder group analysis in Uluabat Lake, Turkey. 2003 Environ. Manage. 31(4), 518–531.
\bibitem{b11} Özesmi, U., Özesmi, S.L.: Ecological models based on people knowledge: A multy – step fuzzy cognitive mapping approach. 2004 Ecol. Model. 176, 43–64.
\bibitem{b12} Misthos, L.S., Messaris, G., Damigos, D., Menegaki, M.: Exploring the perceived intrusion of mining into the landscape using the fuzzy cognitive mapping approach. 2017 Ecol. Eng. 101, 60–74.
\bibitem{b13} Jetter, A.: Fuzzy cognitive maps for engineering and technology management: what works in practice? In: 2006 PICMET Proceedings.
\bibitem{b14} Jetter, A., Schweinfort, W.: Building scenarios with fuzzy cognitive maps: an exploratory study of solar energy. 2010 Futures 43, 52–66.
\bibitem{b15} Papageorgiou, E.I., Groumpos, P.P. A weight adaptation method for fuzzy cognitive map learning. 2005 Soft Comput 9, 846–857.
\bibitem{b16} Gray S. A.; Gray, S.; Cox, L. J.;  Henly-Shepard, S. Mental Modeler: A Fuzzy-Logic Cognitive Mapping Modeling Tool for Adaptive Environmental Management, in: Proceedings of 46th Hawaii International Conference on System Sciences 2013, Wailea, Maui, HI, pp. 965-973.
\bibitem{b17} Horx, Matthias. Die Zukunft nach Corona - Wie eine Krise unsere Gesellschaft, unser Denken und unser Handeln verändert, 2020, Econ Verlag, Berlin.
\bibitem{b18} Widmann, Aloysius. Ist die Corona-Pandemie eine Naturkatastrophe? Der Standard 2020, April 20th.
\bibitem{b19} Detting, Daniel. Zukunftsszenarien nach Corona - Gesundheit wird nicht mehr nur eine individuelle Angelegenheit sein, Der Tagesspiegel 2020, April 5th.
\bibitem{b20} von der Gracht, Heiko. Die Zukunft nach Corona, KPMG Klardenker (eds.) 2020, July 2nd.
\bibitem{b21} Thierstein, Alain. Digitale Transformation im urbanen Raum, Stadt Bauwelt 2018, Nr. 219 / 19.2018 – Digitale Stadt, pp. 32-35.
\bibitem{b22} Goger, Gerald; Piskernik, Melanie; Urban, Harald. Studie: Potenziale der Digitalisierung im Bauwesen 2018, Bundesministerium für Verkehr, Innovation und Technologie and Wirtschaftskammer Österreich, Geschäftsstelle Bau, Vienna / Austria.
\bibitem{b23} Roland Berger Holding GmbH. (eds.). Wirtschaftliche Auswirkungen des Coronavirus - Teil 2, Roland Berger 2020, March 26th, München.
\bibitem{b24} Internet Economy Foundation (IE.F); Roland Berger Holding GmbH. (eds.) Deutschland digital - Sieben Schritte in die Zukunft (undated).
\bibitem{b25} Katz, Raul; Koutroumpis, Pantelis; Callorda, Fernando. Using a digitization index to measure the economic and social impact of digital agendas 2014, info. Vol. 16 No. 1, pp. 32-44.
\bibitem{b26} Roland Berger Holding GmbH. (eds.). Digitales Arbeiten in Zeiten von COVID-19, Roland Berger 2020, April 2nd, München.
\bibitem{b27} Schüller, K.; Förster, A. Digital Literacy für die Stadt, Informationen zur Raumentwicklung 1 2017, p. 108-121.
\bibitem{b28} Simondon, Gilbert. Die Existenzweise technischer Objekte 2012, diaphanes, Zürich.
\bibitem{b29} Baumanns, Thomas et al. (2016): Bauwirtschaft im Wandel, Trends und Potenziale bis 2020, Studie Roland Berger GmbH. und UniCredit Bank AG, München.
\bibitem{b30}  Bundeskanzleramt und Bundesministerium für Wissenschaft, Forschung und Wirtschaft (eds.). Digital Roadmap Austria, 2016, December, Vienna, Austria.
\bibitem{b31} BRZ Deutschland GmbH. (eds.). IT-Trends in der Baubranche 2016 - Status quo und Perspektiven, 2016 Nürnberg.
\bibitem{b32} Hacking, Ian. ‘Style’ for historians and philosophers, Studies in History and Philosophy of Science, March 1992, Part A, Volume 23, Issue 1, pp. 1-20.
\bibitem{b33} Fleck, Ludwik. Entstehung und Entwicklung einer wissenschaftlichen Tatsache - Einführung in die Lehre vom Denkstil und Denkkollektiv 1980, Suhrkamp Verlag, Frankfurt
\bibitem{b34} Kuhn, Thomas S. The Structure of Scientific Revolutions, 1962, University of Chicago Press, Chicago
\bibitem{b35} Sciortino, L. On Ian Hacking’s Notion of Style of Reasoning. Erkenntnis - An International Journal of Scientific Philosophy 2017, 82, pp. 243–264.
\bibitem{b36} Foucault, M. The order of things. 1994, Vintage Books, New York.
\bibitem{b37} Lakatos, I. The methodology of scientific research programmes, 1978, Cambridge Univ. Press.
\bibitem{b38} Kühl, S. Leitbilder erarbeiten, Eine kurze organisationstheoretisch informierte Handreichung, Universität Bielefeld, 2017, Springer Fachmedien Wiesbaden.
\bibitem{b39} Weiser, Mark. The Computer of the 21st Century, Scientific American 1991, 265 (3)
\bibitem{b40} Gross, Neil. The Earth Will Don An Electronic Skin, Bloomberg Business, 1999 August 30th.
\bibitem{b41} Wiegerling, Klaus. Ubiquitous Computing als konkrete Utopie, in: Grimm, Petra; Capurro, Rafael (eds.). Informations- und Kommunikationsutopien, 2008, Franz Steiner Verlag, Stuttgart.
\bibitem{b42} Tanner, Ken.  Common Sense: Get It, Use It, and Teach It in the Workplace, Apress Business, 2013.
\bibitem{b43} Kosko; Bart. Fuzzy cognitive maps, in: International Journal of Man-Machine Studies, 1986, Volume 24, Issue 1, Pages 65-75, ISSN 0020-7373.
\bibitem{b44} Axelrod, R.: Structure of Decision: The Cognitive Maps of Political Elites. 1976 Princeton University Press, Princeton.
\bibitem{b45} Caselles, Antonio, An application of fuzzy cognitive maps to improve well‐being, sustainability and the globalization process, in: Systems Research and Behavioral Science, November 2013, Vol.30(6), pp.646-660.
\bibitem{b46} Bottero M., Datola G., Monaco R. Exploring the Resilience of Urban Systems Using Fuzzy Cognitive Maps. in: Gervasi O. et al. (eds) Computational Science and Its Applications – ICCSA 2017, 2017, Lecture Notes in Computer Science, vol 10406. Springer, Cham.
\bibitem{b47} Cloud, David J., Applied Modeling and Simulation, 1998 McGraw-Hill, New York.
\bibitem{b48} Felix, G., Nápoles, G., Falcon, R. et al. A review on methods and software for fuzzy cognitive maps. Artif Intell Rev 2019, 52, 1707–1737.
\bibitem{b49} Box, George E. P., Empirical Model-Building and Response Surfaces, 1987 Wiley.
\bibitem{b50} G{\"o}deritz, Johannes; Rainer, Roland; Hoffmann, Hubert. Die gegliederte und aufgelockerte Stadt, 1957, T{\"u}bingen. \\
\end{thebibliography}
\end{document}